\documentclass[12pt]{article}
\usepackage{amsmath,amssymb}
\usepackage[mathscr]{euscript}
\usepackage{color}
\usepackage{epsfig}
\usepackage[all]{xypic}

\newtheorem{prop}{Proposition}[section]

\newtheorem{lemm}{Lemma}[section]
\newtheorem{theo}{Theorem}[section]
\newtheorem{coro}{Corollary}[section]

\newcommand{\bbox}{\normalsize {}%
        \nolinebreak \hfill $\blacksquare$ \medbreak \par}

\newcommand{\iN}{\hbox{ {\leaders\hrule\hskip.2cm}{\vrule height
      .22cm} }}

\newcommand{\bfsigma}{\boldsymbol{\sigma}}
\newcommand{\bfeta}{\boldsymbol{\eta}}
\newcommand{\bfA}{\boldsymbol{A}}
\newcommand{\bfU}{\boldsymbol{U}}
\newcommand{\bfV}{\boldsymbol{V}}
\newcommand{\bfH}{\boldsymbol{H}}
\newcommand{\bfp}{\boldsymbol{p}}
\newcommand{\bfF}{\boldsymbol{F}}
\newcommand{\bff}{\boldsymbol{f}}
\newcommand{\bfh}{\boldsymbol{h}}
\newcommand{\bfvarphi}{\boldsymbol{\varphi}}
\newcommand{\bfpsi}{\boldsymbol{\psi}}
\newcommand{\bfgamma}{\boldsymbol{\gamma}}
\newcommand{\bfGamma}{\boldsymbol{\Gamma}}
\newcommand{\bfepsilon}{\boldsymbol{\epsilon}}
\newcommand{\bfchi}{\boldsymbol{\chi}}
\newcommand{\bfPhi}{\boldsymbol{\Phi}}
\newcommand{\bfu}{\boldsymbol{u}}

\newcommand{\g}{\mathfrak{g}}
\newcommand{\G}{\mathfrak{G}}

\newcommand{\R}{\mathbb{R}}

\usepackage{stmaryrd}                       % package pour les intervalles d'entiers...
\title{Multisymplectic formulation of Yang--Mills equations
and Ehresmann connections}
\author{Fr{\'e}d{\'e}ric \textsc{H{\'e}lein}\footnote{Institut de Math{\'e}matiques de Jussieu--Paris Rive Gauche,
UMR CNRS 7586 Universit{\'e} Denis Diderot --- Paris 7,
UFR de Math{\'e}matiques, Case 7012, B{\^a}timent Sophie Germain
75205 Paris Cedex 13, France, \textsf{helein@math.jussieu.fr}}
}

\begin{document}
\maketitle

\textbf{Abstract} --- \emph{We present a multisymplectic formulation of the Yang--Mills equations.
The connections are represented by normalized equivariant 1-forms on
the total space of a principal bundle, with values in a Lie algebra.
Within the multisymplectic framework we realize that, under reasonable hypotheses,
it is not necessary to assume the equivariance condition a priori, since this condition is
a consequence of the dynamical equations.}\\

The motivation of the following work was at first to provide a Hamiltonian formulation of the Yang--Mills system of equations
which would be as much covariant as possible. This means that we look for a formulation which does not depend on choices
of space-time coordinates nor on the trivialization of the principal bundle. Among all possible frameworks
(covariant phase space, etc.)
we favor the multisymplectic formalism which takes automatically into account the locality of fields theories.
Following this approach we have been led to discover a new variational formulation of the Yang--Mills equations
with nice geometrical features.

The origin of the multisymplectic formalism goes back to the discovery by V. Volterra in 1890
\cite{volterra1,volterra1} of generalizations
of the Hamilton equations for variational problems with several variables. These ideas were first developped mainly around
1930 \cite{caratheodory,DeDonder,Weyl,Lepage} and in 1950 \cite{Dedecker}. After 1968 this theory was
geometrized in a way analogous to the construction of symplectic geometry by several mathematical physicists
\cite{garcia,goldstern,krupka} and in particular by a group 
around W. Tulczyjew in Warsaw \cite{Kijowski1,Kijowski2,KijowskiSzczyrba}. This theory has many recent developments
which we cannot report here (see e.g. \cite{GIMMSY,hk1,hk2,Rovelli,Baez,forgergomes,lopez-marsden,kanatchikov1,kanatchikov2} and,
for surveys, \cite{snyatycki,CantrijnIdL,forgerromero,helein,heleinleeds}). Today the Hamilton--Volterra equations are often
called the De Donder--Weyl equations for reference to \cite{DeDonder,Weyl}, which is inaccurate \cite{heleinleeds}. 
However in this paper we name them the
HVDW equations for Hamilton--Volterra--De Donder--Weyl.

The basic concept is the notion of a multisymplectic 
$(n+1)$-form $\omega$ on a smooth manifold $\mathcal{N}$, where $n$ refers to the number of independent variables. The form $\omega$ is 
always \emph{closed} and one often assumes that it is non degenerate, i.e. that the only vector
field $\xi$ on the manifold such that $\xi\iN \omega = 0$ is zero.
An extra ingredient is a Hamiltonian function $H:\mathcal{N}\longrightarrow \R$. One can then
describe the solutions of the HVDW equations by oriented $n$-dimensional
submanifolds $\Gamma$ of $\mathcal{N}$ which satisfy the condition that, at any point
$\textsc{m}\in \mathcal{N}$, there exists a basis $(X_1,\cdots,X_n)$ of $T_\textsc{m}\Gamma$
such that $X_1\wedge \cdots \wedge X_n\iN \omega = (-1)^ndH$.
Equivalently one can replace $\omega$ by its restriction to the level set $H^{-1}(0)$
and describe the solutions as the submanifolds $\Gamma$ of $H^{-1}(0)$ such that
$X_1\wedge \cdots \wedge X_n\iN \omega = 0$ everywhere (plus some independence conditions,
see e.g. \cite{heleinleeds}).

All that have led to elegant formulations of most variational
problems in mathematical physics involving e.g. maps and sections of bundles.
However the multisymplectic formulation of the Yang--Mills raises difficulties \cite{kanatchikov1.5},
because the dynamical field is a connection and is subject to gauge invariance,
hence their geometrical description is delicate. 
A possible approach consists in building a suitable reduction of
the geometry of connections on a $\G$-principal bundle as for instance
in \cite{BrunoCV}. We follow another approach, which is based
on ideas which are now quite standard since {\'E}lie Cartan: we lift the connection defined
on some manifold $\mathcal{M}$ representing the space-time to the principal bundle $\mathcal{P}$ over $\mathcal{M}$ with
structure group $\G$. The connection is then represented
 by a $\g$-valued 1-form $\bfeta$ on $\mathcal{P}$ which satisfies a \emph{normalization} (\ref{normalization})
 and an \emph{equivariance} (\ref{equivariance-a-droite})
 hypothesis. Although a priori necessary the equivariance condition has the drawback of being a constraint on
 the first order derivatives of the field, which, to our opinion, is not a natural condition. 
 
 In the following we compute the Legendre transform for the Yang--Mills action 
 by treating connections as normalized and equivariant $\g$-valued 1-forms on $\mathcal{P}$.
 We find that the natural multisymplectic manifold can be built from the vector
 bundles $\g\otimes T^*\mathcal{P}$ and $\g^*\otimes \Lambda^{n+r-2}T^*\mathcal{P}$ over $\mathcal{P}$, where
 $n+r$ is the dimension of $\mathcal{P}$,
 $\g$ is the  structure Lie algebra and $\g^*$ its dual vector space. These vector
 bundles are endowed with a canonical $\g$-valued 1-form $\eta$ and a canonical
 $\g^*$-valued $(n+r-2)$-form $p$ respectively. Inside $\g\otimes T^*\mathcal{P}$ we consider the
 subbundle $\g\otimes^\textsc{n} T^*\mathcal{P}$ of \emph{normalized} forms. Then the multisymplectic
 manifold corresponds more or less to the total space of the vector bundle
 $\R\oplus_\mathcal{P} (\g\otimes^\textsc{n} T^*\mathcal{P})\oplus_\mathcal{P}(g^*\otimes \Lambda^{n+r-2}T^*\mathcal{P})$,
 equipped with the $(n+r)$-form $\theta = \epsilon \beta\wedge \gamma + p\wedge (d\eta +\eta\wedge \eta)$,
 where $\epsilon$ a coordinate on $\R$ and $\beta\wedge \gamma$ is the volume form on $\mathcal{P}$.
 The Hamiltonian function $H$ is (up to a factor $-\frac{1}{4}$) the squared norm of the coefficients
 $p^{\mu\nu}$ such that $p\wedge dx^\mu\wedge dx^\nu + p^{\mu\nu}\beta\wedge \gamma=0$.
 Once this is done, we will see that we may \textbf{remove} the unnatural equivariance
 constraint and derive the corresponding generalized Hamilton equations without this
 assumption; then we discover that, if the structure group of the gauge theory
 is unimodular and compact (which is the case for $U(1)$ and all $SU(k)$'s),
 the dynamical Hamilton equations \emph{force the $\g$-valued 1-forms to be equivariant}
 and give us back the Yang--Mills equations.
 
 What are the byproducts of this approach ? The fact that we obtain a first order formulation of the
 Yang--Mills equations is not new. But, most importantly, this formulation works
 on the space of normalized $\g$-valued 1-forms
 on a principal bundle which are not equivariant, i.e. which don't correspond to a connection
 in the usual sense. Instead these 1-forms correspond to Ehresmann connections
 on the total space of the bundle $\mathcal{P}$. However the classical Euler--Lagrange equations contains conditions which,
 under some hypotheses on the structure group, forces
 these fields to be equivariant on-shell and hence to correspond to a connection, which turns
 out to be also a solution of the Yang--Mills equation. Hence, although it is different from
 the standard Yang--Mills variational problem, this problem has the same classical
 solutions. We also note that
 our problem is invariant under an action of the standard gauge group of the usual Yang--Mills action,
 plus the action of another gauge group, which is Abelian and acts additively on the momentum
 variables.
 
 It is interesting to note that our new Lagrangian density in (\ref{new-Lagrangian}) is not that
 mysterious and could have been invented out of the blue. The merit of the multisymplectic approach
 here is to provide a conceptual way to build it from the standard Yang--Mills action. In
 particular, performing the Legendre transform \emph{by respecting the equivariance constraint}
 produces automatically the extra fields $p^{aj}$ which play the role of Lagrange multipliers for this
 constraint. A more miraculous fact is however that these extra fields which may not be equivariant
 themselves are dynamically decoupled from the other fields if the gauge structure group is compact unimodular.
 
 Various interesting questions can be set. It seems interesting to study the quantization
 of this model and, in particular, to explore the mass gap question \cite{kanatchikov1.5} in this setting.
 Indeed one could expect that the elastic mechanism which replaces the usual equivariance constraint
 could induce a mass at the quantum level.
 Another point is that our formulation has some flavor of
 a Kaluza--Klein theory, so it would be interesting to study gravitational theories by
 following similar ideas and to build a Kaluza--Klein 
 gravitational theory
 where the mechanism of spontaneaous dynamical reduction that we observed here could be useful.\\
 
 \noindent
 \emph{Acknowledgement} --- The Author is pleased to thank Igor Kanatchikov for useful comments.

 \section{Geometric preliminaries}
 
 \subsection{Yang--Mills gauge fields}\label{section-ym-gauge-fields}
 We are interested in the critical points of a Yang--Mills action functional on an
 $n$-dimensional manifold $\mathcal{M}$ with coordinates $(x^1,\cdots,x^n)$.
 We fix some Lie group $\G$, which will be the structure group of our gauge theory,
 and we denote by $\g$ its Lie algebra.
 The fields are then $\g$-valued 1-forms $\bfA = \bfA_\mu dx^\mu$ on $\mathcal{M}$. The curvature
 2-form of $\bfA$ is $\bfF = d\bfA + \frac{1}{2}[\bfA\wedge \bfA]$. We will assume for simplicity that $\G$
 is a group of matrices and write equivalently $\bfF = d\bfA + \bfA\wedge \bfA$. We have
in local coordinates $\bfF = \frac{1}{2}\bfF_{\mu\nu}dx^\mu\wedge dx^\nu$, where
 $\bfF_{\mu\nu}:= \frac{\partial \bfA_\nu}{\partial x^\mu} - \frac{\partial \bfA_\mu}{\partial x^\nu}
 + [\bfA_\mu,\bfA_\nu]$. We fix a pseudo-Riemannian metric $\textsf{g}_{\mu\nu}$ on $\mathcal{M}$ and a
 metric $\textsf{h}_{ij}$ on $\g$ which is invariant under the adjoint action of $\G$.
 Then the Yang--Mills action of $\bfA$ is
 \begin{equation}\label{standard-ym}
  \mathcal{YM}[\bfA]:= \int_\mathcal{M} d\hbox{vol}_\textsf{g}\left(-\frac{1}{4}|\bfF|^2\right)
 \end{equation}
 where $|\bfF|^2 = \textsf{g}^{\lambda\nu}(x)\textsf{g}^{\mu\sigma}(x)\textsf{h}_{ij}\bfF^i_{\lambda\mu}\bfF^j_{\nu\sigma}$
 and $d\hbox{vol}_\textsf{g}$ is the pseudo-Riemannian measure on $\mathcal{M}$. This action is invariant by gauge
 transformations $\bfA\longmapsto f^{-1}df + f^{-1}\bfA f$, for any map $f$ from $\mathcal{M}$ to $\G$.
 It is well-known that one interprets geometrically a gauge field $\bfA$ as a connection
 on a principal bundle with structure group $\G$ over $\mathcal{M}$.
 Our first task will be to recast this problem by replacing the gauge fields $\bfA$ by
 $\g$-valued 1-forms defined on the total space of the principal bundle, which satisfy
 suitable conditions.

 \subsection{Working on the principal bundle}
 Let $\mathcal{P}$ be the total space of a principal bundle which
 is fibered over $\mathcal{M}$ and with structure group $\G$. We denote by $\pi_\mathcal{M}:\mathcal{P}\longrightarrow \mathcal{M}$ the fibration map.
 We assume that $\G$ is acting on the right on $\mathcal{P}$:
\[
 \begin{array}{ccc}
  \mathcal{P}\times \G & \longrightarrow & \mathcal{P}\\
(z,g) & \longmapsto & z\cdot g = R_g(z)
 \end{array}
\]
This induces an infinitesimal action of $\g$: to any $\xi\in \g$,
we associate the vector field $\rho_\xi$ on $\mathcal{P}$ defined by:
$\forall z\in \mathcal{P}$, $\forall \xi\in \g$, 
$\rho_\xi(z):=  \frac{d}{dt}(z\cdot e^{t\xi})|_{t=0}$;
we also set $\rho_\xi(z) = z\cdot \xi$.
For any $z\in \mathcal{P}$ the orbit of the $\G$ action containing $z$ is the fiber $\mathcal{P}_x$,
where $x=\pi_\mathcal{M}(z)$; the tangent vector subspace to $\mathcal{P}_x$ at $z$ is called the \emph{vertical} subspace and
is denoted by $V_z := \hbox{ker}d(\pi_\mathcal{M})_z$. Since the map $\G\ni g\longmapsto z\cdot g\in \mathcal{P}_x$
is a diffeomorphism, $V_z$ is isomorphic to the Lie
algebra $\g$ of $\G$ through the differential of the latter diffeomorphism: 
\[
 \begin{array}{ccc}
  \mathfrak{g} & \longrightarrow & T_z\mathcal{P}_x = V_z\\
\xi & \longmapsto & z\cdot \xi
 \end{array}
\]
We denote by $\alpha_z:V_z\longrightarrow \mathfrak{g}$ the inverse map. Then $\alpha|_{\mathcal{P}_x}$ is a $\g$-valued
1-form on $\mathcal{P}_x$ {(the Maurer--Cartan form) and is characterized by one of the two following
conditions: $\forall z\in \mathcal{P}_x$,
\begin{equation}\label{definition-maurer-cartan-droite}
 \left[z\cdot \alpha_z(v) = v,\quad \forall v\in V_z\right]\quad
\Longleftrightarrow \quad
\left[\alpha_z(z\cdot \xi) = \xi,\quad \forall \xi\in \mathfrak{g}\right].
\end{equation}

An \emph{Ehresmann connection} on $\mathcal{P}$ is a  distribution of `horizontal' vector subspaces $(H_z)_{z\in \mathcal{P}}$, where
 $\forall z\in \mathcal{P}$, $H_z\subset T_z\mathcal{P}$ and:
\[
 \forall z \in \mathcal{P},\quad H_z\oplus V_z = T_z\mathcal{P}.
\]
Ehresmann connections can be described by using the space $\Gamma(\mathcal{P},\g\otimes T^*\mathcal{P})$ of
sections of the bundle $\g\otimes T^*\mathcal{P}$ over $\mathcal{P}$,
i.e. of $\g$-valued 1-forms on $\mathcal{P}$. Indeed any Ehresmann connection $(H_z)_{z\in \mathcal{P}}$
can be defined by some $\bfeta\in \Gamma(\mathcal{P},\g\otimes T^*\mathcal{P})$
such that $\hbox{ker}\bfeta_z = H_z$, $\forall z\in \mathcal{P}$. The 1-form $\bfeta$ is unique if, furthermore,
it satisfies the \emph{normalization} condition
\begin{equation}\label{normalization}
 \bfeta_z|_{V_z} = \alpha_z,\quad \forall z\in \mathcal{P}.
\end{equation}
We denote by $\Gamma_\textsc{n}(\mathcal{P},\g\otimes T^*\mathcal{P})$ the subspace of $\bfeta\in \Gamma(\mathcal{P},\g\otimes T^*\mathcal{P})$
which satisfy (\ref{normalization}). Alternatively we define the `normalized' 
affine subbundle of the bundle $\g\otimes T^*\mathcal{P}$ to be:
\[
 \g \otimes^\textsc{n}T^*\mathcal{P}:= \{(z,\eta)\in \g\otimes T^*\mathcal{P}; \forall \xi\in \g,
 \eta(z\cdot\xi) = \xi\}
\]
and observe that $\Gamma_\textsc{n}(\mathcal{P},\g\otimes T^*\mathcal{P})$ is the space of sections of $\g \otimes^\textsc{n}T^*\mathcal{P}$.

Among the forms in $\Gamma_\textsc{n}(\mathcal{P},\g\otimes T^*\mathcal{P})$ the ones which lift gauge fields in the sense of
Paragraph \ref{section-ym-gauge-fields}
are characterized by the additional \emph{equivariance} condition:
\begin{equation}\label{equivariance-a-droite}
 \forall (g,z)\in \G\times \mathcal{P}, 
\quad \left(R_g^*\bfeta\right)_z = Ad_{g^{-1}}\circ \bfeta_z = g^{-1}\bfeta_z g,
\end{equation}
where $R_g^*$ is the pull-back by the right action mapping $R_g$. 
We denote by $\Gamma_\textsc{n}^\g(\mathcal{P},\g\otimes T^*\mathcal{P})$ the subspace of \emph{normalized} and
\emph{equivariant} $\g$-valued 1-forms on $\mathcal{P}$.
Assuming that
$\G$ is connected, Condition (\ref{equivariance-a-droite})
is equivalent to its Lie algebraic analogue:
\begin{equation}\label{equivariance-lie-algebraic}
 L_{\rho_\xi}\bfeta + [\xi,\bfeta] = 0,\quad \forall \xi\in\g,
\end{equation}
where $L_{\rho_\xi}$ is the Lie derivative. Lastly
if $\bfeta\in \Gamma_\textsc{n}^\g(\mathcal{P},\g\otimes T^*\mathcal{P})$ the quantity $d\bfeta +\bfeta\wedge \bfeta$ represents
the \emph{curvature} of the connection defined by $\bfeta$ on $\mathcal{M}$.

All that is made clear through a trivialization of $\mathcal{P}$.
Let $\bfsigma: \mathcal{M}\longrightarrow \mathcal{P}$ be a section of $\mathcal{P}$. Then
\[
 \begin{array}{ccc}
  \mathcal{M}\times \G & \longrightarrow & \mathcal{P}\\
(x,g) & \longmapsto & \bfsigma(x)\cdot g
 \end{array}
\]
is a local diffeomorphism. Its inverse map:
\[
 \begin{array}{cccc}
  \varphi:& \mathcal{P} & \longrightarrow & \mathcal{M}\times \G\\
& z & \longmapsto & (x,g)
 \end{array}
\quad\hbox{where}\quad x = \pi_\mathcal{M}(z) \hbox{ and }\bfsigma(x)\cdot g = z,
\]
provides us with a coordinate system. In this setting (\ref{definition-maurer-cartan-droite})
reads $\alpha|_{\mathcal{P}_x} = g^{-1}dg$. Using local coordinates $(x^1,\cdots,x^n)$ on $\mathcal{M}$
and the identification $\bfeta \simeq \varphi^*\bfeta$, we can translate the normalization
condition (\ref{normalization}) by:
\begin{equation}\label{normalization-in-trivialisation}
 \bfeta_{(x,g)} = g^{-1}dg + \bfeta_\mu(x,g)dx^\mu.
\end{equation}
Setting $\bfA_\mu(x,g):= g\bfeta_\mu(x,g)g^{-1}$ and $\bfA_{(x,g)}:= \bfA_\mu(x,g)dx^\mu$,
(\ref{normalization-in-trivialisation}) reads
\begin{equation}\label{variant-normalization-in-trivialisation}
 \bfeta_{(x,g)} = g^{-1}dg +g^{-1}\bfA_{(x,g)} g.
\end{equation}
We then have the identity
\begin{equation}%\label{curvature-P-to-M}
 d\bfeta + \bfeta\wedge \bfeta = g^{-1}(d\bfA+\bfA\wedge \bfA)g.
\end{equation}
Still assuming (\ref{normalization-in-trivialisation}) the extra equivariance
condition (\ref{equivariance-a-droite}) then translates as $\bfA_\mu(x,g) = \bfA_\mu(x)$,
i.e. that $\bfA_\mu$ does not depend on $g\in \G$. If so,
\begin{equation}\label{connexion-dans-trivialisation-droite}
 \bfeta_{(x,g)} = g^{-1}dg +g^{-1}\bfA_x g,
\quad \hbox{where}\quad \bfA_x:= \bfA_\mu(x)dx^\mu.
\end{equation}
In particular the pull-back of $\bfeta$ by $\bfsigma$ is $\bfsigma^*\bfeta = \bfA$ and,
if $\bfsigma':\mathcal{M}\longrightarrow \mathcal{P}$ is another section, then
there exists $\bfgamma:\mathcal{M}\longrightarrow \G$ such that $\bfsigma'(x) = \bfsigma(x)\cdot\bfgamma(x)$,
$\forall x\in \mathcal{M}$ and the pull-back of $\bfeta$ by $\bfsigma'$ is:
$(\bfsigma')^*\bfeta = \bfgamma^{-1}d\bfgamma  + \bfgamma^{-1} \bfA \bfgamma$.
This shows the correspondence between the normalized and equivariant $\g$-valued 1-forms
on $\mathcal{P}$ on the one hand, and the connection 1-forms on the corresponding principal bundle
up to gauge transformations on the other hand.

\subsection{Coframe on the total space $\mathcal{P}$}
Let $(t_1,\cdots, t_r)$ be a basis of $\g$ and, for $i=1,\cdots, r$, set $\rho_i:= \rho_{t_i}$.
We hence obtain a rank $r$ family of tangent vector fields on $\mathcal{P}$ which, at every point
$z\in \mathcal{P}$, spans the vertical subspace $V_z$.
We also choose a local orthonormal moving frame $(\underline{e}_1,\cdots,\underline{e}_n)$
on $\mathcal{M}$. This means that we are given a reference pseudo-Riemannian metric $\textsf{h}_{ab}$
with constant coefficients on $\R^n$ and
that $\langle \underline{e}_a,\underline{e}_b\rangle = \textsf{h}_{ab}$, $\forall a,b=1,\cdots,n$.
In order to obtain a moving frame on $\mathcal{P}$ we choose a section $\bfsigma:\mathcal{M}\longrightarrow \mathcal{P}$ which induces
a trivialization $z = \bfsigma(x)\cdot g \simeq (x,g)$ and we set
\[
 e_a(z):= d(R_g\circ \bfsigma)_x(\underline{e}_a(x))\simeq 
 \underline{e}_a(x)\cdot g,
 \quad \hbox{for }a = 1,\cdots, n.
\]
Then $(e_1,\cdots ,e_n,\rho_1,\cdots, \rho_r)$ is a moving frame on $\mathcal{P}$. We
define its dual coframe
\[
(\beta^1,\cdots,\beta^n,\gamma^1,\cdots,\gamma^r),
\]
i.e. the family
of sections of $T^*\mathcal{P}$ such that $\beta^a(e_b) = \delta^a_b$, $\gamma^i(\rho_j) = \delta^i_j$
and $\beta^a(\rho_j) = \gamma^i(e_b) = 0$.
This provides us with coordinates on $\g\otimes T^*\mathcal{P}$: a point $(z,\eta)$ in $\g\otimes T^*\mathcal{P}$
(where $z\in \mathcal{P}$ and $\eta\in \g\otimes T^*_z\mathcal{P}$) has
the coordinates $(x,g,\eta^i_a,\eta^i_j)$, where $z=\bfsigma(x)\cdot g$ and
$\eta = t_i(\eta^i_a\beta^a + \eta^i_j\gamma^j)$.

Let $\underline{\nabla}$ be the Levi-Civita connection on $T\mathcal{M}$ for the metric
$\textsf{g}_{\mu\nu}$ on $\mathcal{M}$ and let $\omega^b_a \in \Omega^1(\mathcal{M})$ be the connection 1-forms of $\underline{\nabla}$
in the moving frame $(\underline{e}_1,\cdots,\underline{e}_n)$, i.e. such that
$\underline{\nabla}_{\underline{e}_a}\underline{e}_b = \omega^c_b(\underline{e}_a)\underline{e}_c$.
Using $\underline{\nabla}$ and the the choice of a section $\bfsigma$, we construct 
a connection $\nabla$ on $T\mathcal{P}$ : we extend $\omega^b_a$ on $\mathcal{P}$ by letting 
$\omega^b_a\simeq (\pi_\mathcal{M})^*\omega^b_a$ and we set
\[
 \begin{array}{cccccccc}
  \nabla_{e_a}e_b & = & \omega^c_b(e_a)e_c & ; & \nabla_{\rho_i}e_b & = & 0 &;\\
  \nabla_{e_a}\rho_j & = & 0 & ; & \nabla_{\rho_i}\rho_j & = & 0 & .
 \end{array}
\]
This connection acts on sections $\bfeta$ of $\Gamma(\mathcal{P},\g\otimes T^*\mathcal{P})$: if 
$\bfeta = \bfeta_a\beta^a + \bfeta_i\gamma^i$, where $\bfeta_a$ and $\bfeta_i$ are functions
on $\mathcal{P}$ with values in $\g$, then $\forall v\in T_z\mathcal{P}$,
$\nabla_v \bfeta = \left(d\bfeta_a(v) - \omega^b_a(v)\bfeta_b\right)\beta^a + d\bfeta_i(v)\gamma^i$.
Moreover, because of the torsion free conditions
\begin{equation}\label{torsionfree}
d\beta^a + \omega^a_b\wedge\beta^b = 0,
\end{equation}
we have the following expression for the exterior differential of
$\bfeta$,
\begin{equation}\label{deta-covariant}
 d\bfeta = d\bfeta_a\wedge\beta^a - \bfeta_c\omega^c_b\wedge\beta^b
 + d\bfeta_i\wedge \gamma^i + \bfeta_id\gamma^i.
\end{equation}
Hence in particular the $\beta^a\wedge\beta^b$ component of the curvature
$d\bfeta + \bfeta\wedge \bfeta$ is
\begin{equation}\label{composante-courbure-in-frame}
 \bfF_{ab} = (d\bfeta_b - \bfeta_c\omega^c_b)(e_a) -
 (d\bfeta_a - \bfeta_c\omega^c_a)(e_b) + [\bfeta_a,\bfeta_b] =
 (\nabla_{e_a}\bfeta)_b - (\nabla_{e_b}\bfeta)_a + [\bfeta_a,\bfeta_b].
\end{equation}

In the decomposition $\bfeta = \bfeta_a\beta^a + \bfeta_i\gamma^i$ of some $\bfeta\in
\Gamma(\mathcal{P},\g\otimes T^*\mathcal{P})$,
conditions (\ref{normalization}) and (\ref{equivariance-lie-algebraic}) can be expressed
as follows. The normalization condition
means that $\bfeta_i = t_i$, so that (\ref{normalization})  reads
\begin{equation}\label{normalization-inframe}
 \bfeta = \bfeta_a\beta^a + t_i\gamma^i
\end{equation}
and, since $L_{\rho_\xi}\beta^a =0$, $\forall \xi\in\g$,  the equivariance condition (\ref{equivariance-lie-algebraic}) reads
\begin{equation}\label{equivariance-inframe}
 d\bfeta_a(\rho_i) + [t_i,\bfeta_a] = 0, \quad \forall i = 1,\cdots, r.
\end{equation}

Let us denote by $c_{ij}^k$ the constants such that
\begin{equation}\label{def-structure-constants}
 [t_i,t_j]= c_{ij}^kt_k,
\end{equation}
where the summation over repeated indices is assumed. Then from the decomposition
$g^{-1}dg = t_i\gamma^i$ and the zero curvature condition $d(g^{-1}dg) + (g^{-1}dg)\wedge (g^{-1}dg) = 0$
we deduce the relation
\begin{equation}\label{structure-dgamma}
 d\gamma^i + \frac{1}{2}c_{jk}^i\gamma^j\wedge \gamma^k = 0.
\end{equation}

To conclude, we define $\beta:= \beta^1\wedge \cdots \wedge \beta^n$
and $\gamma = \gamma^1\wedge \cdots \wedge \gamma^r$ and,
for $1\leq a,b\leq n$, $1\leq i,j\leq r$, we set (the symbol $\iN$ denotes the interior product)
\[
 \beta_a:= e_a\iN \beta,\quad \beta_{ab}:= e_b\iN (e_a\iN \beta),
 \quad \gamma_i:= \rho_i\iN \gamma,\quad \gamma_{ij}:= \rho_j\iN (\rho_j\iN \gamma).
\]
We note the following useful relations
\begin{equation}\label{betayoga}
\beta^a\wedge \beta_b = \delta^a_b \beta,\quad
\beta^a\wedge \beta^b \wedge \beta_{cd} = (\delta^a_c\delta^b_d 
- \delta^a_d\delta^b_c)\beta
\end{equation}
and similarly
\begin{equation}\label{gammayoga}
\gamma^i\wedge \gamma_j = \delta^i_j \gamma,\quad
\gamma^i\wedge \gamma^j \wedge \gamma_{kl} = (\delta^i_k\delta^j_l
- \delta^i_l\delta^j_k)\gamma.
\end{equation}
The following result will be helpful later on. We recall that, if $\xi\in \g$,
then $\hbox{ad}_\xi:\g\longrightarrow\g$ is the linear map defined by $\hbox{ad}_\xi(\eta) = [\xi,\eta]$,
$\forall \eta\in \g$.
\begin{lemm} The following identities holds
\begin{enumerate}
 \item For any $i = 1, \cdots ,r$,
 \begin{equation}\label{identity-dgamma-i}
  d\gamma_i + \hbox{tr}(\hbox{ad}_{t_i})\gamma = 0.
 \end{equation}
 \item For any $a,b = 1,\cdots, n$,
 \begin{equation}\label{identity-dbeta-a}
  d\beta_a = \omega^b_a\wedge\beta_b,
 \end{equation}
 \begin{equation}\label{identity-dbeta-ab}
  d\beta_{ab} = \omega^c_a\wedge \beta_{cb} + \omega^c_b\wedge\omega_{ac}.
 \end{equation}
\end{enumerate}
\end{lemm}
\emph{Proof} --- The proof of (\ref{identity-dgamma-i}) follows from the following computation, where we assume 
a summation over each repeated index and we use (\ref{structure-dgamma}) and (\ref{gammayoga}),
\[
  d\gamma_i = d\gamma^j\wedge \gamma_{ij}
  = - \frac{1}{2}c^j_{kl}\gamma^k\wedge \gamma^l\wedge \gamma_{ij}
  = - c^j_{ij}\gamma = -\hbox{tr}(\hbox{ad}_{t_i})\gamma;
\]
(\ref{identity-dbeta-a}) and (\ref{identity-dbeta-ab}) are obtained by similar computations, by
using (\ref{torsionfree}) and $\omega^a_b+\omega^b_a=0$:
\[
 d\beta_a = d\beta^b\wedge\beta_{ab} = - \omega^b_c\wedge\beta^c\wedge \beta_{ab}
 = - \omega^b_b\wedge\beta_a + \omega^b_a\wedge\beta_b,
\]
\[
 d\beta_{ab} = d\beta^c\wedge \beta_{abc} = -\omega^c_d\wedge\beta^d\wedge\beta_{abc}
 = - \omega^c_c\wedge\beta_{ac} + \omega^c_b\wedge\beta_{ac} - \omega^c_a\wedge\beta_{bc}.
\]
\hfill $\square$\\
Identity (\ref{identity-dgamma-i}) has the following straightforward consequence.
We recall that a Lie algebra is unimodular iff
$\hbox{tr}(\hbox{ad}_\xi) = 0$, $\forall \xi\in \g$. Note that $U(1)$ and all $SU(k)$'s are unimodular.
\begin{coro}\label{corollary-unimodular}
 Assume that $\g$ is unimodular, then $d\gamma_i = 0$, $\forall i=1,\cdots ,r$.
\end{coro}

\section{Towards the multisymplectic formulation}
\subsection{The multisymplectic framework}
In order to set the multisymplectic framework 
it is simpler to start with an abstract general description: let $\mathcal{Z}$ be a $m$-dimensional
manifold and consider the fiber bundle $\Lambda^mT^*\mathcal{Z}$ of $m$-forms over $\mathcal{Z}$. By using the
fibration $\pi_\mathcal{Z}:\Lambda^mT^*\mathcal{\mathcal{Z}} \longrightarrow \mathcal{Z}$ we define a canonical $m$-form
$\theta^\mathcal{Z}$ on $\Lambda^mT^*\mathcal{Z}$ by $\theta^\mathcal{Z}_{(Z,\varpi)}(X_1,\cdots ,X_m)
:= \varpi(\pi_\mathcal{Z}^*X_1,\cdots, \pi_\mathcal{Z}^*X_m)$, $\forall Z\in \mathcal{Z}$,
$\forall \varpi\in \Lambda^mT^*_Z\mathcal{Z}$, $\forall X_1,\cdots, X_m\in T_{(Z,\varpi)}(\Lambda^mT^*\mathcal{Z})$.
If $\mathcal{Z}$ is itself fibered over a manifold $\mathcal{X}$ by a projection map $\pi_\mathcal{X}:\mathcal{Z}
\longrightarrow \mathcal{X}$, this defines in each tangent space
$T_Z\mathcal{Z}$ a vertical subspace $V_Z$ which is the kernel of $\pi_\mathcal{X}^*$. We then define
the subbundle of so-called $(m-1)$-horizontal forms (see \cite{forgergomes})
\[
 \Lambda^m_1T^*\mathcal{\mathcal{Z}}:=
 \{(\varpi,Z)\in \Lambda^mT^*\mathcal{\mathcal{Z}}; \forall v_1,v_2\in V_Z,v_1\wedge v_2\iN \varpi = 0\}.
\]
This corresponds to assuming that each $m$-multilinear map $\varpi\in \Lambda^m_1T^*_Z\mathcal{\mathcal{Z}}$
has a degree at most one in the vertical coordinates of vectors in $T_Z\mathcal{Z}$. Then
$\Lambda^m_1T^*\mathcal{\mathcal{Z}}$ is the geometrical framework for the so-called
`De Donder--Weyl' theory for sections of $\mathcal{Z}$ over $\mathcal{X}$
which are critical points of a first order variational problem \cite{hk1}. We will denote
by $\theta^\mathcal{Z}_1$ the restriction of $\theta^\mathcal{Z}$ to  $\Lambda^m_1T^*\mathcal{\mathcal{Z}}$.

We use this setting for $m=n+r$, $\mathcal{Z} = \g\otimes T^*\mathcal{P}$ and $\mathcal{X} = \mathcal{P}$.
Coordinate functions on $\Lambda^m_1T^*(\g\otimes T^*\mathcal{P})$ are $(x^\mu,g)$ for a point $z\in \mathcal{P}$,
$(\eta^i_a,\eta^i_j)$ for the components of $\eta \in \g\otimes T_z^*\mathcal{P}$
in the basis $(t_i\otimes \beta^a, t_i\otimes \gamma^j)$ and
$(e,p^{ab}_i,p^{jb}_i,p^{aj}_i,p^{jk}_i)$ for the components of
$\varpi\in \Lambda^m_1T^*_{(z,\eta)}(\g\otimes T^*\mathcal{P})$ in the basis
$(\beta\wedge \gamma, d\eta^i_a \wedge \beta_b\wedge \gamma,
d\eta^i_j\wedge \beta_b\wedge \gamma, (-1)^nd\eta^i_a\wedge \beta\wedge \gamma_j,
(-1)^nd\eta^i_j\wedge \beta\wedge \gamma_k)$.
The Poincar{\'e}--Cartan
form $\theta^\mathcal{Z}_1$ then reads
\begin{equation}\label{theta-volterra-general}
\begin{array}{cccccc}
 \theta^\mathcal{Z}_1 = & e\beta\wedge \gamma  & + & p^{ab}_id\eta^i_a \wedge \beta_b\wedge \gamma
 & + & p^{jb}_id\eta^i_j\wedge \beta_b\wedge \gamma \\
 & & + & (-1)^np^{a j}_id\eta^i_a\wedge \beta\wedge \gamma_j
 & + &  (-1)^np^{jk}_id\eta^i_j\wedge \beta\wedge \gamma_k.
 \end{array}
\end{equation}

Since we are interested in normalized sections of $\g\otimes T^*\mathcal{P}$,
i.e. satisfying (\ref{normalization}), we must actually work on
$\Lambda^{n+r}_1T^*(\g\otimes^\textsc{n} T^*\mathcal{P})$. The latter space is a bundle over 
$\g\otimes^\textsc{n} T^*\mathcal{P}$ and can actually
be constructed through a reduction of $\Lambda^{n+r}_1T^*(\g\otimes T^*\mathcal{P})$:
we restrict ourself on $(\pi_{\g\otimes T^*\mathcal{P}})^{-1}(\g\otimes^\textsc{n} T^*\mathcal{P})$
and for any $(z,\eta)\in \g\otimes^\textsc{n} T^*\mathcal{P}$, we replace the fiber
$\Lambda^{n+r}_1T^*_{(z,\eta)}(\g\otimes T^*\mathcal{P})$ by its quotient
by the annihilator of $T_{(z,\eta)}(\g\otimes^\textsc{n} T^*\mathcal{P})$, i.e. the space of
forms $\varpi$ in $\Lambda^{n+r}_1T^*_{(z,\eta)}(\g\otimes T^*\mathcal{P})$
such that $v\iN \varpi =0$, $\forall v\in T_{(z,\eta)}(\g\otimes^\textsc{n} T^*\mathcal{P})$.
This amounts to impose (see also (\ref{normalization-inframe}))
\begin{equation}\label{impose-zetaij}
 \eta^i_j = \delta^i_j
\end{equation}
and to assume that $(\tilde{e},\tilde{p}^{ab}_i,\tilde{p}^{a j}_i,
\tilde{p}^{jb}_i,\tilde{p}^{jk}_i) \sim (e,p^{ab}_i, p^{a j}_i,
p^{jb}_i,p^{jk}_i)$ whenever $(\tilde{e},\tilde{p}^{ab}_i,\tilde{p}^{a j}_i)
= (e,p^{ab}_i, p^{a j}_i)$, so that we may forget about coordinates
$(p^{jb}_i,p^{jk}_i)$.
Denoting simply by $\theta$ the restriction to $\Lambda^{n+r}_1T^*(\g\otimes^\textsc{n} T^*\mathcal{P})$ of $\theta^\mathcal{Z}_1$
given in (\ref{theta-volterra-general}),
this leads to the simplification
\begin{equation}\label{theta-simple}
 \theta =  e\beta\wedge \gamma  + p^{ab}_id\eta^i_a \wedge \beta_b\wedge \gamma
   + (-1)^np^{a j}_id\eta^i_a\wedge \beta\wedge \gamma_j.
\end{equation}

\subsection{The Legendre correspondence}
A Lagrangian for a gauge theory is a real valued function $L$ defined on the
bundle $T^*\mathcal{P}\otimes_{\g\otimes^\textsc{n} T^*\mathcal{P}} (T(\g\otimes^\textsc{n} T^*\mathcal{P})/T\mathcal{P})$ over
$\g\otimes^\textsc{n} T^*\mathcal{P}$, whose fiber at $(z,\eta)\in \g\otimes^\textsc{n} T^*\mathcal{P}$ is
the space of linear maps $\lambda:T_z\mathcal{P}\longrightarrow T_{(z,\eta)}(\g\otimes^\textsc{n} T^*\mathcal{P})$
such that $d(\pi_\mathcal{P})_{(z,\eta)}\circ \lambda = \hbox{Id}_{T_z\mathcal{P}}$
(this vector space can be canonically identified with
$T^*\mathcal{P}_z\otimes (T_{(z,\eta)}(\g\otimes^\textsc{n} T^*\mathcal{P})/T_z\mathcal{P})$).
We define coordinates $(x,g,\eta^i_a,\lambda^i_{a;b},\lambda^i_{a;j})$
on $T^*\mathcal{P}\otimes_{\g\otimes^\textsc{n} T^*\mathcal{P}} (T(\g\otimes^\textsc{n} T^*\mathcal{P})/T\mathcal{P})$
in a natural way from the ones on $\g\otimes T^*\mathcal{P}$:
for any $(z,\eta,\lambda)\in T^*\mathcal{P}\otimes_{\g\otimes^\textsc{n} T^*\mathcal{P}}
(T(\g\otimes^\textsc{n} T^*\mathcal{P})/T\mathcal{P})$, take a section $\bfeta\in \Gamma(\mathcal{P},\g\otimes^\textsc{n} T^*\mathcal{P})$
such that $\bfeta(z) = \eta$ and (viewing $\bfeta$ as a map from $\mathcal{P}$ to the total space
of the bundle $\g\otimes^\textsc{n} T^*\mathcal{P}$) the differential of $\bfeta$ at $z$ is $\lambda$.
Then $\lambda$ has the coordinates $\lambda^i_{a;b}(z,\eta,\lambda):= 
(\nabla_{e_b}\bfeta^i)_a = (d(\bfeta^i_a)_z - \bfeta^i_c\omega^c_a)(e_b)$
and $\lambda^i_{a;j}(z,\eta,\lambda):= (\nabla_{\rho_j}\bfeta^i)_a=  d(\bfeta^i_a)_z(\rho_j)$.

However we have to take into account the following important fact. The problem we start with
concerns gauge fields on a space-time manifold $\mathcal{M}$ but \emph{not all} normalized $\g$-valued 1-forms $\bfeta$ on $\mathcal{P}$,
so that we actually need to compute the Legendre correspondence along \emph{equivariant}
1-forms $\bfeta$. In view of (\ref{equivariance-inframe}) this means that we must impose the
extra constraint on $\lambda$
\begin{equation}\label{extra-constraint}
 \lambda^i_{a;j} = [\eta_a,t_j]^i.
\end{equation}
We denote by $T^*\mathcal{P}\otimes_{\g\otimes^\textsc{n} T^*\mathcal{P}} (T(\g\otimes^\textsc{n} T^*\mathcal{P})/T\mathcal{P})^\g$
the submanifold of points $(z,\eta,\lambda)\in
T^*\mathcal{P}\otimes_{\g\otimes^\textsc{n} T^*\mathcal{P}} (T(\g\otimes^\textsc{n} T^*\mathcal{P})/T\mathcal{P})$ which
satisfy Condition (\ref{extra-constraint}).

The standard Yang--Mills Lagrangian in
(\ref{standard-ym}) has the following expression by using the moving
frame $(e_a,\rho_i)$:
\begin{equation}\label{standard-ym-preLegendre}
 L(z,\eta,\lambda) = -\frac{1}{4} \textsf{g}^{ac}\textsf{g}^{bd}\textsf{h}_{ij}F^i_{ab}F^j_{cd},
\end{equation}
where (see (\ref{composante-courbure-in-frame}))
\[
 F^i_{ab} = \lambda^i_{b;a} - \lambda^i_{a;b} 
 + [\eta_a,\eta_b]^i.
\]
Such a Lagrangian induces a correspondence
between $T^*\mathcal{P}\otimes_{\g\otimes^\textsc{n} T^*\mathcal{P}} (T(\g\otimes^\textsc{n} T^*\mathcal{P})/T\mathcal{P})^\g$
and a submanifold of $\Lambda^{n+r}_1T^*(\g\otimes^\textsc{n} T^*\mathcal{P})$
as follows (see \cite{hk1}).

Assume as in the previous section that the coframe $(\beta^a,\gamma^i)$ is orthonormal, 
then the volume element $d\hbox{vol}_\textsf{g}$ in (\ref{standard-ym}) is equal to $\beta\wedge \gamma$.
We define the function $W$ on
$\left(T^*\mathcal{P}\otimes_{\g\otimes^\textsc{n} T^*\mathcal{P}} (T(\g\otimes^\textsc{n} T^*\mathcal{P})/T\mathcal{P})^\g\right)
\times_{\g\otimes^\textsc{n} T^*\mathcal{P}} \Lambda^{n+r}_1T^*(\g\otimes^\textsc{n} T^*\mathcal{P})$
(sorry for the  notation) by:
\[
 W(z,\eta,\lambda,\varpi):= \theta_{(z,\eta,\varpi)}(\lambda(e_1),\cdots, \lambda(e_n),
 \lambda(\rho_1),\cdots, \lambda(\rho_r)) - L(z,\eta,\lambda)
\]
and we say that \emph{$(z,\eta,\lambda)$ is in correspondence with $(z,\eta,\varpi)$ if
$\frac{\partial W}{\partial \lambda}(z,\eta,\lambda,\varpi)= 0$}. (This condition
amounts to say that $(z,\eta,\lambda,\varpi)$ is a critical point of the restriction of
$W$ to the fiber of the map $(z,\eta,\lambda,\varpi) \longmapsto (z,\eta,\varpi)$.)
If so the value
of $W$ at $(z,\eta,\lambda,\varpi)$ defines a function $H$ of $(z,\eta,\varpi)$, which is
the Hamiltonian.

We now need to compute
$\theta_{(z,\eta,\varpi)}(\lambda(e_1),\cdots, \lambda(e_n),
\lambda(\rho_1),\cdots, \lambda(\rho_r))$. In order to avoid a
messy computation we use the following trick: choose the right coframe (as we learned from Cartan).
Here given some $(z,\eta,\lambda,\varpi)$,
we replace the coframe $(\beta^a,\gamma^i,d\eta^i_a)$ by $(\beta^a,\gamma^i,\delta\eta^i_a)$
in the expression of $\theta_{(z,\eta,\varpi)}$, where
\[
 \delta\eta^i_a = d\eta^i_a - \lambda^i_{a;b}\beta^b - \eta^i_c\omega^c_a
 - \lambda^i_{a;j}\gamma^j,
\]
so that
\begin{equation}\label{reason-of-the-trick}
 \forall v\in T_z\mathcal{P},\quad
 \delta\eta^i_a(\lambda(v)) = 0.
\end{equation}
Hence by using (\ref{extra-constraint}),
\begin{equation}\label{def-deltazeta}
 d\eta^i_a = \delta\eta^i_a + \lambda^i_{a;b}\beta^b + \eta^i_c\omega^c_a
 + [\eta_a,t_j]^i\gamma^j
\end{equation}
This gives us by using (\ref{betayoga}) and (\ref{gammayoga})
\[
\begin{array}{ccccl}
 \theta & = & e\beta\wedge \gamma  & + & p^{ab}_i(\delta\eta^i_a 
+ \lambda^i_{a;c}\beta^c + \eta^i_c\omega^c_a + [\eta_a,t_k]^i \gamma^k)
\wedge \beta_b\wedge \gamma\\
& & & + & (-1)^np^{a j}_i(\delta\eta^i_a 
+ \lambda^i_{a;c}\beta^c + \eta^i_c\omega^c_a + [\eta_a,t_k]^i\gamma^k)
 \wedge \beta\wedge \gamma_j
 \end{array}
 \]
and, noting $\Gamma^c_{ab}:= \omega^c_b(e_a)$ (so that $\omega^c_b = \Gamma^c_{ab}\beta^a$), 
\[
\begin{array}{ccccl}
 & = & e\beta\wedge \gamma  & + &  p^{ab}_i(\lambda^i_{a;b} + \eta^i_c\Gamma^c_{ba})\beta\wedge \gamma
 + p^{a j}_i[\eta_a,t_j]^i \beta\wedge \gamma \\
 & & & + &p^{ab}_i\delta\eta^i_a \wedge \beta_b\wedge \gamma
 + (-1)^np^{a j}_i\delta\eta^i_a \wedge \beta\wedge \gamma_j.
 \end{array}
\]
Hence by using (\ref{reason-of-the-trick}) it follows that
\[
 \theta_{(z,\eta,\varpi)}(\lambda(e_1),\cdots, \lambda(e_n),
\lambda(\rho_1),\cdots, \lambda(\rho_r)) =
e + p^{ab}_i(\lambda^i_{a;b} + \eta^i_c\Gamma^c_{ba})
 + p^{a j}_i[\eta_a,t_k]^i
\]
and thus
\[
 W(z,\eta,\lambda,\varpi) = e + p^{ab}_i(\lambda^i_{a;b} + \eta^i_c\Gamma^c_{ba})
 + p^{a j}_i[\eta_a,t_j]^i - L(z,\eta,\lambda).
\]
We hence find immediately that the condition $\frac{\partial W}{\partial \lambda} =0$
reads
\begin{equation}\label{Legendre-relations}
 p^{ab}_i = \frac{\partial L}{\partial \lambda^i_{a;b}}(z,\eta,\lambda).
\end{equation}
We apply this relation with the standard Yang--Mills action (\ref{standard-ym-preLegendre}) and
find
\begin{equation}\label{p-par-Legendre}
 p^{ab}_i = \textsf{h}_{ij}\textsf{g}^{ac}\textsf{g}^{bd}F^j_{cd}
 =  \textsf{h}_{ij}\textsf{g}^{ac}\textsf{g}^{bd}\left(\lambda^j_{d;c}
 - \lambda^j_{c;d} + [\eta_c,\eta_d]^j\right).
\end{equation}
We observe that $\varpi$ is subject to the constraints
\begin{equation}\label{firstclass}
 p^{ab}_i + p^{ba}_i = 0.
\end{equation}
We thus define the image of the Legendre correspondence:
\[
 \mathcal{N}:= \{(z,\eta,\varpi)\in \Lambda^{n+r}_1T^*(\g\otimes^\textsc{n} T^*\mathcal{P});
 p^{ab}_i + p^{ba}_i = 0\}.
\]
We still denote by $\theta$ the restriction of $\theta$ on $\mathcal{N}$ and set $\omega:= d\theta$:
$(\mathcal{N},\omega)$ is the multisymplectic manifold we will work with.
Assuming (\ref{p-par-Legendre}) we deduce from (\ref{standard-ym-preLegendre}) that
$L(z,\eta,\lambda) = - \frac{1}{4}\textsf{h}^{ij}\textsf{g}_{ac}\textsf{g}_{bd}p^{ab}_ip^{cd}_j$
and, by using (\ref{firstclass}),
\[
 \begin{array}{ccl}
  p^{ab}_i(\lambda^i_{a;b} + \eta^i_c\Gamma^c_{ba}) & = &
  - \frac{1}{2}p^{ab}_i(\lambda^i_{b;a} + \eta^i_c\Gamma^c_{ab} - \lambda^i_{a;b} - \eta^i_c\Gamma^c_{ba})\\
 & = & - \frac{1}{2}p^{ab}_i(\lambda^i_{b;a} - \lambda^i_{a;b} + [\eta_a,\eta_b]^i) 
%  \\ & & \quad
  - \frac{1}{2}p^{ab}_i\eta^i_c(\Gamma^c_{ab} - \Gamma^c_{ba}) + \frac{1}{2}p^{ab}_i[\eta_a,\eta_b]^i\\
  & = & - \frac{1}{2}p^{ab}_iF^i_{ab} 
  - \frac{1}{2}p^{ab}_i\eta^i_c(\Gamma^c_{ab} - \Gamma^c_{ba}) + \frac{1}{2}p^{ab}_i[\eta_a,\eta_b]^i\\
 & = & - \frac{1}{2}p^{ab}_i\textsf{h}^{ij}\textsf{g}_{ac}\textsf{g}_{bd}p^{bd}_i
  - \frac{1}{2}p^{ab}_i\eta^i_c(\Gamma^c_{ab}- \Gamma^c_{ba})
   + \frac{1}{2}p^{ab}_i[\eta_a,\eta_b]^i
 \end{array}
\]
We hence deduce the expression for the Hamiltonian function $H$
\begin{equation}
 H(z,\eta,\varpi) = e - \frac{1}{4}\textsf{h}^{ij}\textsf{g}_{ac}\textsf{g}_{bd}p^{ab}_ip^{cd}_j
 - \frac{1}{2}p^{ab}_i\eta^i_c(\Gamma^c_{ab}- \Gamma^c_{ba})
 + \frac{1}{2}p^{ab}_i[\eta_a,\eta_b]^i + p^{a j}_i[\eta_a,t_j]^i.
\end{equation}

\subsection{Change of coordinates}
We will change the coordinates on $\mathcal{N}$
in order to simplify the expression of the Hamiltonian function and in such a way that $\theta$
depends on $\eta$ uniquely through the quantity $d\eta + \eta\wedge \eta$.
We set
\[
 \epsilon:= e - \frac{1}{2}p^{ab}_i\eta^i_c(\Gamma^c_{ab}- \Gamma^c_{ba})
 + \frac{1}{2}p^{ab}_i[\eta_a,\eta_b]^i + p^{a j}_i[\eta_a,t_j]^i.
\]
We note then that 
\begin{equation}
 H(z,\eta,\varpi) = \epsilon - \frac{1}{4}\textsf{h}^{ij}\textsf{g}_{ac}\textsf{g}_{bd}p^{ab}_ip^{cd}_j.
\end{equation}
Moreover, from (\ref{theta-simple}),
\begin{equation}\label{theta-provisoire}
\begin{array}{ccccl}
 \theta & = & \epsilon & + & \displaystyle  p^{ab}_i\left(d\eta^i_a \wedge \beta_b\wedge \gamma
 + \frac{1}{2}\eta^i_c(\Gamma^c_{ab}- \Gamma^c_{ba}) \beta\wedge\gamma
 - \frac{1}{2}[\eta_a,\eta_b]^i\beta\wedge\gamma\right)\\
& & & + & \displaystyle  p^{a j}_i\left((-1)^nd\eta^i_a\wedge \beta\wedge \gamma_j
   - [\eta_a,t_j]^i\beta\wedge\gamma\right).
\end{array}
\end{equation}
In order to transform this expression, we need some preliminaries.
First by setting $\eta_a = t_i\eta^i_a$ and 
$\eta = \eta_a\beta^a + t_i\gamma_i$ (the canonical $\g$-valued 1-form on
$\g\otimes^\textsc{n}T^*\mathcal{P}$), we get by using (\ref{deta-covariant}):
\[
 d\eta = d\eta_a\wedge \beta^a - \eta_c\omega^c_b\wedge\beta^b
 - \frac{1}{2}[t_j,t_k]\gamma^j\wedge \gamma^k.
\]
Since on the other hand
\[
 \eta\wedge \eta = \frac{1}{2}[\eta_a,\eta_b]\beta^a\wedge\beta^b +
 \frac{1}{2}[t_i,t_j]\gamma^i\wedge\gamma^j + [\eta_a,t_i]\beta^a\wedge\gamma^i,
\]
we deduce that
\begin{equation}\label{decomposition-dzeta+}
 d\eta + \eta\wedge \eta = d\eta_a\wedge \beta^a
 + \left(\frac{1}{2}[\eta_a,\eta_b] -\eta_c\Gamma^c_{ab}\right)\beta^a\wedge\beta^b 
 + [\eta_a,t_i]\beta^a\wedge\gamma^i.
\end{equation}
This implies by using (\ref{betayoga}) that
\begin{equation}\label{from-betayoga}
 \left(d\eta + \eta\wedge \eta\right)\wedge \beta_{ab}\wedge\gamma =
 \left(d\eta_b\wedge\beta_a - d\eta_a\wedge\beta_b
 - \eta_c(\Gamma^c_{ab}- \Gamma^c_{ba}) \beta
 + [\eta_a,\eta_b]\beta\right) \wedge \gamma
\end{equation}
and by using (\ref{gammayoga}):
\begin{equation}\label{from-gammayoga}
 \left(d\eta + \eta\wedge \eta\right)\wedge \beta_a\wedge\gamma_j =
 (-1)^n\left((-1)^nd\eta_a\wedge\beta\wedge\gamma_j 
 + [t_j,\eta_a]\beta\wedge \gamma\right).
\end{equation}
Hence we deduce from (\ref{from-betayoga}) and (\ref{firstclass})
the second r.h.s. term of (\ref{theta-provisoire}):
\begin{equation}\label{partie-p-lambdamu}
 p^{ab}_i\left(d\eta^i_a \wedge \beta_b
 + \frac{1}{2}\eta^i_c(\Gamma^c_{ab}- \Gamma^c_{ba}) \beta
 - \frac{1}{2}[\eta_a,\eta_b]^i\beta\right)\wedge \gamma
 = - \frac{1}{2}p^{ab}_i
 \left(d\eta + \eta\wedge \eta\right)^i\wedge \beta_{ab}\wedge\gamma
\end{equation}
and from (\ref{from-gammayoga}) the last r.h.s. term of (\ref{theta-provisoire}):
\begin{equation}\label{partie-p-lambdaj}
 p^{a j}_i\left((-1)^nd\eta^i_a\wedge \beta\wedge \gamma_j
   - [\eta_a,t_j]^i\beta\wedge\gamma\right) =
   (-1)^np^{a j}_i(d\eta + \eta\wedge\eta)^i\wedge\beta_a\wedge\gamma_j.
\end{equation}
We thus deduce by summarizing (\ref{theta-provisoire}), (\ref{partie-p-lambdamu}) and (\ref{partie-p-lambdaj}):
\begin{prop}
 The Poincar{\'e}--Cartan form $\theta$ on $\mathcal{N}$ reads:
 \begin{equation}\label{poincare-cartan-1}
  \theta = \epsilon \beta\wedge\gamma + p_i\wedge (d\eta + \eta\wedge\eta)^i,
 \end{equation}
where
\begin{equation}\label{def-pi-dual}
 p_i:= - \frac{1}{2}p^{ab}_i\beta_{ab}\wedge\gamma 
 + (-1)^np^{a j}_i\beta_a\wedge\gamma_j.
\end{equation}
\end{prop}
An alternative expression is $\theta = \epsilon \beta\wedge\gamma + p\wedge (d\eta + \eta\wedge\eta)$,
where in the r.h.s a duality pairing between the $\g^*$-valued coefficients of $p$
and the $\g$-valued coefficients of $d\eta + \eta\wedge \eta$ is assumed.

\subsection{Re-interpretation of the previous result}\label{reinterpretation}
Let us rephrase the previous result. We see a posteriori that the
multisymplectic manifold ($\mathcal{N},\omega)$, where $\omega = d\theta$ and $\theta$ is
given by (\ref{poincare-cartan-1}) and (\ref{def-pi-dual}), has a
simple alternative construction. We consider the pair of vector bundles
$\g\otimes^\textsc{n}T^*\mathcal{P}$ and $\g^*\otimes \Lambda^{n+r-2}T^*\mathcal{P}$ over $\mathcal{P}$ (where $\g^*$ is the
dual vector space of $\g$) and their fibered direct sum over $\mathcal{P}$ with $\R$:
\[
 \widetilde{\mathcal{N}}:= \R\oplus_\mathcal{P}
 \left(\g\otimes^\textsc{n}T^*\mathcal{P}\right)\oplus_\mathcal{P}\left(\g^*\otimes \Lambda^{n+r-2}T^*\mathcal{P}\right).
\]
The base $\mathcal{P}$ is equipped with the volume form $\beta\wedge\gamma$ and 
$\epsilon$ is a coordinate on $\R$.
Denote by $(p^{ab},p^{aj},p^{jk})$ the $\g^*$-valued coordinates on the fibers of $\g^*\otimes \Lambda^{n+r-2}T^*\mathcal{P}$
in the basis $(-\beta_{ab}\wedge \gamma, (-1)^n\beta_a\wedge \gamma_j,\beta\wedge \gamma_{jk})$.
The bundle $\g\otimes^\textsc{n}T^*\mathcal{P}$ is equipped with the canonical 
$\g$-valued 1-form $\eta$ (which reads $\eta_a\beta^a + t_i\gamma^i$ in $\g$-valued coordinates)
and $\g^*\otimes \Lambda^{n+r-2}T^*\mathcal{P}$ with the canonical $\g^*$-valued
$(n+r-2)$-form $p$ (which reads $-\frac{1}{2}p^{ab}\beta_{ab}\wedge\gamma 
 + (-1)^np^{a j}\beta_a\wedge\gamma_j + \frac{1}{2}p^{jk}\beta\wedge \gamma_{jk}$ in
$\g^*$-valued coordinates). We also define the vector subbundles
$\g^*\otimes \Lambda^{n+r-2}_0T^*\mathcal{P}:= \{(x,p\in \g^*\otimes \Lambda^{n+r-2}T^*\mathcal{P};
\forall a,\beta^a\wedge p =0\}$ and
$\g^*\otimes \Lambda^{n+r-2}_1T^*\mathcal{P}:= \{(x,p\in \g^*\otimes \Lambda^{n+r-2}T^*\mathcal{P};
\forall a,b,\beta^a\wedge \beta^b\wedge p =0\}$. In coordinates $\g^*\otimes \Lambda^{n+r-2}_0T^*\mathcal{P}$
is defined by the equations $p^{ab} = p^{aj} = 0$ and  $\g^*\otimes \Lambda^{n+r-2}_1T^*\mathcal{P}$
by $p^{ab} = 0$. We have the obvious inclusions
\[
 \g^*\otimes \Lambda^{n+r-2}_0T^*\mathcal{P}\ \subset \
 \g^*\otimes \Lambda^{n+r-2}_1T^*\mathcal{P}\ \subset \
 \g^*\otimes \Lambda^{n+r-2}T^*\mathcal{P}.
\]
By setting $\theta:= \epsilon \beta\wedge \gamma +
p_i\wedge (d\eta + \eta\wedge \eta)^i$, we obtain the same expression
as (\ref{poincare-cartan-1}), because, in view of (\ref{decomposition-dzeta+}),
all terms involving $p^{jk}_i$ cancel. Hence $(\mathcal{N},\theta)$ is recovered by quotienting out
$\g^*\otimes \Lambda^{n+r-2}T^*\mathcal{P}$ by $\g^*\otimes \Lambda^{n+r-2}_0T^*\mathcal{P}$.

In this setting the Hamiltonian function $H$ has also an intrinsic characterization:
up to a factor $-\frac{1}{4}$, it is the squared norm of all quantities $p^{ab}$
such that $p^{ab}\beta\wedge\gamma + \beta^a\wedge \beta^b \wedge p = 0$.

\section{The HVDW equations}

The multisymplectic form $\omega = d\theta$ on $\mathcal{N}$ is
\begin{equation}\label{multisymplectic-form}
 \omega = d\epsilon\wedge \beta\wedge\gamma
 + dp_i\wedge (d\eta +\eta\wedge\eta)^i +
 (d\eta\wedge \eta - \eta\wedge d\eta)^i\wedge p_i. 
\end{equation}

\subsection{What do we want to do and how to proceed ?}\label{howto}
The geometrical expression of the \emph{HVDW equations} in $(\mathcal{N},\omega)$ for the Hamiltonian
function $H$ consists in a condition on an oriented submanifold $\bfGamma$ of $\mathcal{N}$
of dimension $n+r$ (representing the graph of a solution), which says that, for any point
$\textsc{m}$ of coordinates $(x,g,\eta^i_a,p^{ab}_i,p^{aj}_i)$
of $\bfGamma$,
if $(X_1,\cdots,X_n,Y_1,\cdots,Y_r)$ is a basis of the tangent space to $\bfGamma$ at $\textsc{m}$
such that $\beta\wedge\gamma(X_1,\cdots,X_n,Y_1,\cdots,Y_r) = 1$, then
\begin{equation}\label{volterra-hamilton-basic}
 (X_1\wedge\cdots\wedge X_n\wedge Y_1\wedge\cdots\wedge Y_r)\iN\omega = (-1)^{n+r}dH
\end{equation}
(see \cite{hk1}).
However for the Yang--Mills problem we started from a variational problem on \emph{equivariant}
$\g$-valued 1-forms. But is is not clear a priori whether we should impose a similar
constraint in the Hamiltonian version. In the following we will derive the HVDW
equations in the most general case, i.e. without assuming any equivariance hypothesis a priori.
The HVDW equations with an equivariance constraint will be simply obtained by
adding this extra constraint to the dynamical equations. We will see however that both
approaches work and that, under some reasonable hypotheses, they lead to the Yang--Mills system.

Any fixed $(n+r)$-dimensional submanifold $\bfGamma$ which is a graph can be represented as the image of
an unique embedding of $\mathcal{P}$
in $\R\oplus_\mathcal{P} \left(\g\otimes^\textsc{n}T^*\mathcal{P}\right)\oplus_\mathcal{P}\left(\g^*\otimes \Lambda^{n+r-2}T^*\mathcal{P}\right)$
of the form $\bfu:z\longmapsto (z,\bfepsilon(z),\bfeta(z),\bfp(z))$.
It suffices to estimate the l.h.s. of (\ref{volterra-hamilton-basic})
when replacing $(X_1,\cdots,X_n,Y_1,\cdots,Y_r)$ by
$(\bfu_*e_1,\cdots,\bfu_*e_n,\bfu_*\rho_1,\cdots,\bfu_*\rho_r)$.
However a direct computation
of this quantity can be very messy. So again we use the same trick as for the Legendre transform
and, given some point $\textsc{m}$ of $\bfGamma$ of coordinates $(z,\bfepsilon(z),\bfeta(z),\bfp(z))$, we replace
the coframe $(\beta^a,\gamma^i,d\epsilon,d\eta^i_a,dp^{ab}_i,dp^{aj}_i)$ at
$\textsc{m}$ by $(\beta^a,\gamma^i,\delta\epsilon,\delta\eta^i_a,\delta p^{ab}_i,\delta p^{aj}_i)$,
where
\begin{equation}
 \begin{array}{ccl}
  \delta\epsilon & := & d\epsilon - d\bfepsilon(e_a)\beta^a - d\bfepsilon(\rho_j)\gamma^i\\
  \delta\eta^i_a & := & d\eta^i_a - d\bfeta^i_a(e_b)\beta^b - d\bfeta^i_a(\rho_j)\gamma^j\\
  \delta p^{ab}_i & := & dp^{ab}_i - d\bfp^{ab}_i(e_c)\beta^c - d\bfp^{ab}_i(\rho_j)\gamma^j\\
  \delta p^{aj}_i & := & dp^{aj}_i - d\bfp^{aj}_i(e_b)\beta^b - d\bfp^{aj}_i(\rho_j)\gamma^j,
 \end{array}
\end{equation}
It follows in particular that
\begin{equation}\label{cancelation-property}
 \delta\epsilon\circ d\bfu_z = \delta\eta^i_a\circ d\bfu_z 
 = \delta p^{ab}_i\circ d\bfu_z 
 = \delta p^{aj}_i\circ d\bfu_z = 0.
\end{equation}
It will be useful to introduce the covariant derivatives at $\textsc{m}$
$\bfepsilon_{;a}:= \nabla_{e_a}\bfepsilon = d\bfepsilon(e_a)$, 
$\bfepsilon_{;i}:= \nabla_{\rho_i}\bfepsilon = d\bfepsilon(\rho_i)$,
$\bfeta^i_{b;a}  := (\nabla_{e_a}\bfeta^i)_b = (d\bfeta^i_b - \bfeta^i_c\omega^c_b)(e_a)$,
$\bfp^{ab}_{i;c} := (\nabla_{e_c}\bfp_i)^{ab} = (d\bfp^{ab}_i + \bfp^{db}_i\omega^a_d
+ \bfp^{ad}_i\omega^b_d)(e_c)$, etc.,
so that:
\begin{equation}\label{a-substituer}
 \begin{array}{ccl}
  d\epsilon & = & \delta\epsilon + \bfepsilon_{;a}\beta^a + \bfepsilon_{;i}\gamma^i\\
  d\eta^i_a & = & \delta\eta^i_a + \bfeta^i_{a;b}\beta^b + \bfeta^i_c\omega^c_a + \bfeta^i_{a;j}\gamma^j\\
  d p^{ab}_i & = & \delta p^{ab}_i + \bfp_{i;c}^{ab}\beta^c - \bfp^{cb}_i\omega^a_c
  - \bfp^{ac}_i\omega^b_c + \bfp_{i;j}^{ab}\gamma^j\\
  dp^{aj}_i & = & \delta p^{aj}_i + \bfp_{i;b}^{aj}\beta^b - \bfp^{cj}_i\omega^a_c
  + \bfp^{aj}_{i;j}\gamma^j,
 \end{array}
\end{equation}
where we assume implicitly that the symbols in bold characters denotes components
of $\bfu$ at $z$ such that $\bfu(z) = \textsc{m}$.
In the following we evaluate separately the terms in (\ref{multisymplectic-form})
in view of finding the HVDW equations.

\subsection{The computation of $dp_i\wedge(d\eta + \eta\wedge\eta)^i$}
To enlight the notations we drop here the upper indices, coefficients are thus
$\g$-valued.
Substituting the expression for $d\eta_a$ in (\ref{a-substituer}) and using (\ref{torsionfree})
and (\ref{structure-dgamma}) we obtain
\[
 \begin{array}{ccl}
  d\eta & = & d\eta_a\wedge \beta^a + \bfeta_ad\beta^a + t_id\gamma^i\\
  & = & \left(\delta\eta_a + \bfeta_{a;b}\beta^b + \bfeta_c\omega^c_a + \bfeta_{a;j}\gamma^j\right)
  \wedge \beta^a - \bfeta_a\omega^a_b\wedge \beta^b - \frac{1}{2}[t_i,t_j]\gamma^i\wedge\gamma^j
 \end{array}
\]
hence
\begin{equation}\label{deta}
 d\eta = \delta\eta_a\wedge\beta^a + \frac{1}{2}(\bfeta_{b;a}-\bfeta_{a;b})\beta^a\wedge\beta^b
  - \bfeta_{a;j}\beta^a\wedge\gamma^j - \frac{1}{2}[t_i,t_j]\gamma^i\wedge\gamma^j. 
\end{equation}
On the other hand
\[
\begin{array}{ccl}
 \eta\wedge \eta & = & (\bfeta_a\beta^a + t_i\gamma^i)\wedge  (\bfeta_b\beta^b + t_j\gamma^j)\\
 & = & \frac{1}{2}[\bfeta_a,\bfeta_b]\beta^a\wedge\beta^b + [\bfeta_a,t_j]\beta^a\wedge\gamma^j
 + \frac{1}{2}[t_i,t_j]\gamma^i\wedge\gamma^j.
\end{array}
\]
Hence
\begin{equation}\label{deta+eta-square}
\begin{array}{ccl}
 d\eta + \eta\wedge\eta & = & \delta\eta_a\wedge\beta^a +
 \frac{1}{2}(\bfeta_{b;a}-\bfeta_{a;b} + [\bfeta_a,\bfeta_b])\beta^a\wedge\beta^b\\
 &  &  -\; (\bfeta_{a;j} - [\bfeta_a,t_j])\beta^a\wedge\gamma^j
\end{array}
\end{equation}
On the other hand we need also to compute $dp$. We drop the lower indices, so that coefficients
are now $\g^*$-valued. This quantity splits in two terms:
\begin{equation}\label{split-dp}
 dp = -\frac{1}{2}d(p^{ab} \beta_{ab}\wedge\gamma)
 + (-1)^nd(p^{ai}\beta_a\wedge\gamma_i).
\end{equation}
Substituting the expression for $dp^{ab}$
given by (\ref{a-substituer}) and using (\ref{identity-dbeta-ab}) we obtain for the
first r.h.s. term of (\ref{split-dp})
\[
 \begin{array}{ccl}
  d(p^{ab} \beta_{ab}\wedge\gamma) & = & dp^{ab}\wedge\beta_{ab}\wedge\gamma 
  + \bfp^{ab}\omega^c_a\wedge\beta_{cb}\wedge\gamma  + \bfp^{ab}\omega^c_b\wedge\beta_{ac}\wedge\gamma\\
  & = & (\delta p^{ab} + \bfp^{ab}_{;c}\beta^c
  - \bfp^{cb}\omega^a_c - \bfp^{ac}\omega^b_c
  + \bfp^{ab}_{;i}\gamma^i)\wedge\beta_{ab}\wedge\gamma \\
 & &  +\; (\bfp^{cb}\omega^a_c\wedge\beta_{ab}  + \bfp^{ac}\omega^b_c\wedge\beta_{ab})\wedge\gamma\\
 & = & \delta p^{ab}\wedge\beta_{ab}\wedge\gamma
 + (\bfp^{ab}_{;b}\beta_{a} - \bfp^{ab}_{;a}\beta_{b})\wedge\gamma
 \end{array}
\]
and for the second r.h.s. term of (\ref{split-dp}) we substitute the expression for $dp^{ai}$
given by (\ref{a-substituer}) and we use (\ref{identity-dbeta-a}) and
(\ref{identity-dgamma-i})
\[
 \begin{array}{ccl}
  d(p^{ai}\beta_a\wedge\gamma_i) & = & dp^{ai}\wedge \beta_a\wedge\gamma_i
  + \bfp^{ai}\omega^b_a\wedge \beta_b\wedge\gamma_i + (-1)^n\bfp^{ai}\beta_a\wedge \hbox{tr}(\hbox{ad}_{t_i})\gamma\\
  & = & (\delta p^{ai} + \bfp^{ai}_{;b}\beta^b - \bfp^{bi}\omega^a_b + \bfp^{ai}_{;j}\gamma^j)\wedge \beta_a\wedge\gamma_i\\
  & & +\; \bfp^{bi}\omega^a_b\wedge \beta_a\wedge\gamma_i + (-1)^n\hbox{tr}(\hbox{ad}_{t_i})\bfp^{ai}\beta_a\wedge\gamma\\
  & = & \delta p^{ai}\wedge \beta_a\wedge\gamma_i + \bfp^{ai}_{;a}\beta\wedge\gamma_i
  - (-1)^n(\bfp^{ai}_{;i} - \hbox{tr}(\hbox{ad}_{t_i})\bfp^{ai})\beta_a\wedge\gamma
 \end{array}
\]
Hence
\[
 \begin{array}{ccl}
  dp & = & -\frac{1}{2}\delta p^{ab}\wedge\beta_{ab}\wedge\gamma
 - \bfp^{ab}_{;b}\beta_{a}\wedge\gamma\\
 & &  +\; (-1)^n\delta p^{ai}\wedge \beta_a\wedge\gamma_i + (-1)^n\bfp^{ai}_{;a}\beta\wedge\gamma_i
  - (\bfp^{ai}_{;i} - \hbox{tr}(\hbox{ad}_{t_i})\bfp^{ai})\beta_a\wedge\gamma
 \end{array}
\]
or
\begin{equation}\label{dp}
\begin{array}{ccl}
 dp & = & \displaystyle -\frac{1}{2}\delta p^{ab}\wedge\beta_{ab}\wedge\gamma + (-1)^n\delta p^{ai}\wedge \beta_a\wedge\gamma_i\\
& & \displaystyle  -\; (\bfp^{ab}_{;b}+\bfp^{ai}_{;i} - \hbox{tr}(\hbox{ad}_{t_i})\bfp^{ai})\beta_a\wedge\gamma
  + (-1)^n\bfp^{ai}_{;a}\beta\wedge\gamma_i.
  \end{array}
\end{equation}
The last step consists in computing the product $dp_i\wedge (d\eta+\eta\wedge \eta)^i$. For that
purpose we split $dp = \Pi_1 + \Pi_2 + \Pi_3 + \Pi_4$, where
\[
 \begin{array}{cclcccl}
\Pi_1 & := & -\frac{1}{2}\delta p^{ab}\wedge\beta_{ab}\wedge\gamma & ; & 
\Pi_2 & := & (-1)^n\delta p^{ai}\wedge \beta_a\wedge\gamma_i \\
\Pi_3 & := & -(\bfp^{ab}_{;b}+\bfp^{ai}_{;i} - \hbox{tr}(\hbox{ad}_{t_i})\bfp^{ai})\beta_a\wedge\gamma
& ; & \Pi_4 & := & (-1)^n\bfp^{ai}_{;a}\beta\wedge\gamma_i.
 \end{array}
\]
Similarly we split $d\eta+\eta\wedge\eta = H_1 + H_2 + H_3$, where
\[
 \begin{array}{cclcccl}
  H_1 & := & \delta\eta_a\wedge\beta^a & ; &
  H_2 & := & \frac{1}{2}(\bfeta_{b;a}-\bfeta_{a;b} + [\bfeta_a,\bfeta_b])\beta^a\wedge\beta^b\\
  H_3 & := & -(\bfeta_{a;j} - [\bfeta_a,t_j])\beta^a\wedge\gamma^j. & & & &
 \end{array}
\]
All products $\Pi_J\wedge H_K$ vanish, except the following ones:
\[
\begin{array}{ccl}
 \Pi_1\wedge H_1 & = &  \frac{1}{2}(\delta\eta^i_b\wedge \delta p^{ab}_i\wedge \beta_a - 
 \delta\eta^i_a\wedge \delta p^{ab}_i\wedge \beta_b)\wedge\gamma,\\
 \Pi_2\wedge H_1 & = &  - (-1)^n\delta\eta^i_a\wedge \delta p^{aj}_i\wedge \beta\wedge \gamma_j,\\
 \Pi_3\wedge H_1 & = & 
 - (\bfp^{ab}_{i;b} + \bfp^{aj}_{i;j} - \hbox{tr}(\hbox{ad}_{t_j})\bfp^{aj}) \delta\eta^i_a\wedge \beta\wedge \gamma,\\
 \Pi_1\wedge H_2 & = &  - \frac{1}{2}(\bfeta_{b;a} - \bfeta_{a;b} + [\bfeta_a,\bfeta_b])^i\delta p^{ab}_i \wedge \beta\wedge \gamma,\\
 \Pi_2\wedge H_3 & = &  (\bfeta_{a;j} + [t_j,\bfeta_a])^i\delta p^{aj}_i\wedge \beta\wedge \gamma.
\end{array}
\]
Hence using (\ref{firstclass})
\begin{equation}\label{dpdeta}
\begin{array}{r}
dp_i\wedge (d\eta+\eta\wedge \eta)^i  = \quad
\delta\eta^i_b\wedge \delta p^{ab}_i\wedge \beta_a\wedge\gamma
- (-1)^n\delta\eta^i_a\wedge \delta p^{aj}_i\wedge \beta\wedge \gamma_j\\
 - \left[ (\bfp^{ab}_{i;b} + \bfp^{aj}_{i;j} - \hbox{tr}(\hbox{ad}_{t_j})\bfp^{aj}) \delta\eta^i_a\right.\\
 \left.+ \frac{1}{2}(\bfeta_{b;a} - \bfeta_{a;b} + [\bfeta_a,\bfeta_b])^i\delta p^{ab}_i
 - (\bfeta_{a;j} + [t_j,\bfeta_a])^i\delta p^{aj}_i\right]\wedge \beta\wedge \gamma.
\end{array}
\end{equation}

\subsection{The computation of $(d\eta \wedge \eta - \eta\wedge d\eta)^i\wedge p_i$}
In the following we note $[d\eta\wedge\eta]:= d\eta \wedge \eta - \eta\wedge d\eta$.
From (\ref{deta}) we know that:
\begin{equation}\label{deta-phase1}
 d\eta  = (\delta \eta_a + \bfeta_{a;b}\beta^b + \bfeta_{a;j}\gamma^j) \wedge\beta^a 
 - \frac{1}{2}[t_i,t_j]\gamma^i\wedge \gamma^j.
\end{equation}
On the other hand, we have $[d\eta\wedge\eta] = [d\eta,\bfeta_a]\wedge\beta^a + [d\eta,t_j]\wedge\gamma^j$ and thus
\[
 \begin{array}{ccl}
  [d\eta\wedge\eta]^i\wedge p_i & = & \left([d\eta,\bfeta_b]^i\wedge\beta^b + [d\eta,t_j]^i\wedge\gamma^j\right)
  \wedge \left(-\frac{1}{2}\bfp^{ac}_i\beta_{ac} \wedge\gamma + (-1)^n\bfp^{ak}_i\beta_a\wedge\gamma_k\right)\\
  & = & -\frac{1}{2}[d\eta,\bfeta_b]^i\wedge \bfp^{ac}_i(\delta^b_c\beta_a - \delta^b_a\beta_c)\wedge\gamma\\
  & & + (-1)^n[d\eta,\bfeta_b]^i\wedge \bfp^{ak}_i\delta^b_a\beta\wedge \gamma_k
 + (-1)^n[d\eta,t_j]^i\wedge \bfp^{ak}_i(-1)^{n-1}\delta^j_k\beta_a\wedge \gamma,
 \end{array}
\]
which gives us
\begin{equation}\label{deta-phase2}
 [d\eta\wedge\eta]^i\wedge p_i = - [d\eta,\bfeta_b]^i\wedge \bfp^{ab}_i\beta_a\wedge\gamma
 + (-1)^n[d\eta,\bfeta_a]^i\wedge \bfp^{ak}_i\beta\wedge\gamma_k
 - [d\eta,t_j]^i\wedge \bfp^{aj}_i\beta_a\wedge \gamma.
\end{equation}
The r.h.s. of (\ref{deta-phase2}) is the sum of the three quantities
$M_1:=  - [d\eta,\bfeta_b]^i\wedge \bfp^{ab}_i\beta_a\wedge\gamma$,
$M_2:= (-1)^n[d\eta,\bfeta_a]^i\wedge \bfp^{ak}_i\beta\wedge\gamma_k$ and
$M_3:= - [d\eta,t_j]^i\wedge \bfp^{aj}_i\beta_a\wedge \gamma$.
When substituting the value of $d\eta$ given by (\ref{deta-phase1}) in
(\ref{deta-phase2}), we see that $M_2$ vanishes and we just have
\begin{equation}\label{deta-phase3}
 [d\eta\wedge\eta]^i\wedge p_i =
 \bfp^{ab}_i[\eta_b,\delta\eta_a]^i\wedge\beta\wedge\gamma
 + \bfp^{aj}_i[t_j,\delta\eta_a]^i\wedge\beta\wedge\gamma.
\end{equation}
It is here useful to note that the summation over $i$ of the quantities
$\bfp^{ab}_i[\eta_b,\delta\eta_a]^i$
is a duality product between $\bfp^{ab}\in \g^*$
and $[\bfeta_b,\delta\eta_a] = \hbox{ad}_{\bfeta_b}(\delta\eta_a)\in \g$. It thus coincides with the duality product between
$\hbox{ad}_{\bfeta_b}^*(\bfp^{ab})\in \g^*$ and $\delta\eta_a\in \g$, i.e. with
$\left(\hbox{ad}_{\bfeta_b}^*(\bfp^{ab})\right)_i\delta\eta_a^i$, where $\hbox{ad}_{\bfeta_b}^*$
is the adjoint of $\hbox{ad}_{\bfeta_b}$. Similarly we have
$\bfp^{aj}_i[t_j,\delta\eta_a]^i = \left(\hbox{ad}_{t_j}^*(\bfp^{aj})\right)_i\delta\eta_a^i$.
Hence (\ref{deta-phase3}) reads
\begin{equation}\label{deta-final}
 [d\eta\wedge\eta]^i\wedge p_i =
 \left(\left(\hbox{ad}_{\bfeta_b}^*(\bfp^{ab})\right)_i + \left(\hbox{ad}_{t_j}^*(\bfp^{aj})\right)_i\right)
 \delta\eta_a^i\wedge\beta\wedge\gamma.
\end{equation}

\subsection{Conclusion}
Collecting (\ref{a-substituer}), (\ref{dpdeta}) and (\ref{deta-final}) and substituting
in (\ref{multisymplectic-form}), we obtain
\begin{equation}\label{omega-prepare}
\begin{array}{ccl}
\omega & = &
\delta\eta^i_b\wedge \delta p^{ab}_i\wedge \beta_a\wedge\gamma
- (-1)^n\delta\eta^i_a\wedge \delta p^{aj}_i\wedge \beta\wedge \gamma_j\\
& & +\; \delta \epsilon\wedge \beta\wedge\gamma\\
& & - \left(\bfp^{ab}_{i;b} - \left(\hbox{ad}_{\bfeta_b}^*(\bfp^{ab})\right)_i
+ \bfp^{aj}_{i;j} - \left(\hbox{ad}_{t_j}^*(\bfp^{aj})\right)_i
- \hbox{tr}(\hbox{ad}_{t_j})\bfp^{aj}_i\right) \delta\eta^i_a\wedge \beta\wedge \gamma\\
& & - \frac{1}{2}(\bfeta_{b;a} - \bfeta_{a;b} + [\bfeta_a,\bfeta_b])^i\delta p^{ab}_i\wedge \beta\wedge \gamma\\
& &  + (\bfeta_{a;j} + [t_j,\bfeta_a])^i\delta p^{aj}_i\wedge \beta\wedge \gamma.
\end{array}
\end{equation}
We can now come back to the considerations of Section \ref{howto} and write Equation
(\ref{volterra-hamilton-basic}) with $(X_1,\cdots,X_n,Y_1,\cdots,Y_r)$ equal to
$(\bfu_*e_1,\cdots,\bfu_*e_n,\bfu_*\rho_1,\cdots,\bfu_*\rho_r)$. Writing
$\bfU = X_1\wedge \cdots\wedge X_n\wedge Y_1\wedge \cdots \wedge Y_r$ for short,
we deduce from (\ref{cancelation-property}) that, up to the factor $(-1)^{n+r}$,
the l.h.s. of (\ref{volterra-hamilton-basic})
reduces to:
\begin{equation}\label{XYomega}
 \begin{array}{ccl}
(-1)^{n+r}\bfU\iN \omega & = & \delta \epsilon
- \frac{1}{2}(\bfeta_{b;a} - \bfeta_{a;b} + [\bfeta_a,\bfeta_b])^i\delta p^{ab}_i
+ (\bfeta_{a;j} + [t_j,\bfeta_a])^i\delta p^{aj}_i\\
& & - \left(\bfp^{ab}_{i;b} - \left(\hbox{ad}_{\bfeta_b}^*(\bfp^{ab})\right)_i
+ \bfp^{aj}_{i;j} - \left(\hbox{ad}_{t_j}^*(\bfp^{aj})\right)_i
- \hbox{tr}(\hbox{ad}_{t_j})\bfp^{aj}_i\right) \delta\eta^i_a.
\end{array}
\end{equation}
We observe that
the first line in the r.h.s. of (\ref{omega-prepare}) does not contribute because it contains
terms quadratic in $\delta(\cdot)$.

On the other hand we also need to estimate $dH$. In the following
we use the metric $\textsf{g}_{ab}$ and its inverse $\textsf{g}^{ab}$
to respectively lower and lift indices. We set e.g.
$\bfp_{ab}:= \textsf{g}_{ac}\textsf{g}_{bd}\bfp^{cd}$, etc.
\[
 \begin{array}{ccl}
  dH & = & d\epsilon - \frac{1}{2}\textsf{h}^{ij}p_{abj}dp^{ab}_i\\
  & = & \delta\epsilon - \frac{1}{2}\textsf{h}^{ij}p_{abj}\delta p^{ab}_i
   + \left(\bfepsilon_{;e} -\frac{1}{2}\textsf{h}^{ij}p_{abj} \bfp^{ab}_{i;e}\right)\beta^e
  +  \left(\bfepsilon_{;k} -\frac{1}{2}\textsf{h}^{ij}p_{abj} \bfp^{ab}_{i;k}\right)\gamma^k\\
  & = & \delta\epsilon - \frac{1}{2}\textsf{h}^{ij}p_{abj}\delta p^{ab}_i
  + \bfH_{;e}\beta^e + \bfH_{;k}\gamma^k,
 \end{array}
\]
where we wrote  $\bfH:= \bfepsilon - \frac{1}{4}\textsf{h}^{ij}\bfp_{abj} \bfp^{ab}_i$
for short.

Let us pose $T:= (-1)^{n+r}\bfU\iN\omega - dH$, so that 
(\ref{volterra-hamilton-basic}) reads $T =0$. The previous
computation shows that
\begin{equation}\label{volterra-hamilton-synt}
 \begin{array}{ccl}
T & = &  -\bfH_{;a}\beta^a - \bfH_{;i}\gamma^i \\
 & & - \frac{1}{2}\left(\bfeta_{b;a}^i - \bfeta_{a;b}^i + [\bfeta_a,\bfeta_b]^i
 - \textsf{h}^{ij}\bfp_{abj}\right)\delta p^{ab}_i\\
 & &  + \left(\bfeta_{a;j}^i + [t_j,\bfeta_a]^i\right)\delta p^{aj}_i\\
&  & - \left(\bfp^{ab}_{i;b} - \left(\hbox{ad}_{\eta_b}^*(\bfp^{ab})\right)_i
 + \bfp^{aj}_{i;j} - \left(\hbox{ad}_{t_j}^*(\bfp^{aj})\right)_i
- \hbox{tr}(\hbox{ad}_{t_j})\bfp^{aj}_i \right)\delta \eta^i_a.
 \end{array}
\end{equation}

\section{Classical solutions of the HVDW equations}
We study here the solutions of the HVDW system of equations.
We first note that the vanishing of the coefficients of $\beta^a$ and $\gamma^i$
in (\ref{volterra-hamilton-synt}) means that the solution $\bfGamma$ is contained
in a level set of $H$, a general feature in multisymplectic geometry. In the following
we look more carefully at the other equations.

As a preliminary we introduce some notations. We denote by $\textsf{h}_*:\g\longrightarrow \g^*$
the vector isomorphisme s.t. $(\textsf{h}_*\xi)(\zeta) = \textsf{h}_{ij}\xi^i\zeta^j$, $\forall \xi,\zeta\in\g$
and by $\textsf{h}^*:\g^*\longrightarrow\g$ the inverse mapping. Note that, since the metric
$h$ is invariant by the adjoint action of $\G$ on $\g$, the following relations hold
\begin{equation}\label{adjoint-de-ad}
 \textsf{h}_*[\xi,\zeta] = - \hbox{ad}^*_\xi(\textsf{h}_*\zeta)
 \quad\hbox{and}\quad
 [\xi,\textsf{h}^*\ell] = - \textsf{h}^*(\hbox{ad}_\xi^*\ell),\quad
 \forall \xi,\zeta\in \g,\forall \ell\in \g^*
\end{equation}
%Lastly we set
%\[
% \bfF^{ab} := \textsf{h}^*\left(\bfp^{ab}\right) \quad
% \bfF^{ai} := \textsf{h}^*\left(\bfp^{ai}\right).
%\]

\subsection{The HVDW equations with the equivariance assumption}
We consider here a system of HVDW equations on fields which are
assumed to be \emph{equivariant} a priori. The equivariance condition on $\bfeta$
automatically implies that the coefficients of $\delta p^{aj}_i$ in (\ref{volterra-hamilton-synt})
vanishes. Hence it turns out that the field $p^{aj}_i$ is unuseful and that one
can set it to be equal to zero a priori. This leads to the simplification
\[
 \begin{array}{ccl}
T & = &  -\bfH_{;a}\beta^a - \bfH_{;i}\gamma^i \\
 & & - \frac{1}{2}\left(\bfeta_{b;a}^i - \bfeta_{a;b}^i + [\bfeta_a,\bfeta_b]^i
 - \textsf{h}^{ij}\bfp_{abj}\right)\delta \bfp^{ab}_i\\
&  & - \left(\bfp^{ab}_{i;b} - \left(\hbox{ad}_{\bfeta_b}^*(\bfp^{ab})\right)_i
  \right)\delta \bfeta^i_a.
 \end{array}
\]
Hence equation $T=0$ is equivalent to the condition  
that $H$ is constant along $\bfGamma$ and that $\bfu$ satisfies the system of equations
\[
 \left\{ 
 \begin{array}{rcc}
  \bfeta_{b;a} - \bfeta_{a;b} + [\bfeta_a,\bfeta_b] & = &
  \textsf{h}^*\bfp_{ab}\\
  \bfp^{ab}_{;b} - \hbox{ad}_{\bfeta_b}^*(\bfp^{ab}) & = & 0.
 \end{array}
 \right.
\]
By (\ref{adjoint-de-ad}) one sees that the second equation is equivalent to
$\left(\textsf{h}^*\bfp^{ab}\right)_{;b} +[\bfeta_b,\textsf{h}^*\bfp^{ab}] = 0$, i.e. the Yang--Mills equation.

We note that the first equation implies also
\[
\begin{array}{ccl}
 \bfp^{ab}_{;i} & = & \textsf{h}_*\left(\bfeta_{b;a} - \bfeta_{a;b} + [\bfeta_a,\bfeta_b]\right)_{;i}
 =  - \textsf{h}_*\left(\hbox{ad}_{t_i}\left(\bfeta_{b;a} - \bfeta_{a;b} + [\bfeta_a,\bfeta_b]\right)\right)\\
 & = & \hbox{ad}_{t_i}^*\left(\textsf{h}_*\left(\bfeta_{b;a} - \bfeta_{a;b} + [\bfeta_a,\bfeta_b]\right)\right)
= \hbox{ad}_{t_i}^*\bfp^{ab},
\end{array}
\]
where we used first (\ref{equivariance-inframe}), then (\ref{adjoint-de-ad}). Hence this implies that
$\bfp^{ab}$ is equivariant (we may assume it a priori or not, it does not change the result).

\subsection{The HVDW equations without assuming the equivariance a priori}
Beside the condition that $H$ is constant along a solution $\bfGamma$, the equation $T=0$
gives us the system:
\begin{equation}\label{ham-volt-beau}
 \left\{ 
 \begin{array}{rcc}
  \bfeta_{b;a} - \bfeta_{a;b} + [\bfeta_a,\bfeta_b] & = &
  \textsf{h}^*\bfp_{ab}\\
  \bfeta_{a;j} + [t_j,\bfeta_a] & = & 0\\
  \bfp^{ab}_{;b} - \hbox{ad}_{\bfeta_b}^*(\bfp^{ab})
  + \bfp^{aj}_{;j} - \hbox{ad}_{t_j}^*(\bfp^{aj})
- \hbox{tr}(\hbox{ad}_{t_j})\bfp^{aj} & = & 0.
 \end{array}
 \right.
\end{equation}
\begin{enumerate}
 \item The first equation in (\ref{ham-volt-beau}) is the same as in the previous paragraph.
\item The second equation in (\ref{ham-volt-beau})
is just the equivariance condition (\ref{equivariance-inframe})
for the 1-form $\bfeta$: here this condition is not assumed a priori but is obtained
as one of the dynamical equations ! This is due to
the fact that the fields $\bfp^{aj}_i$ plays the role of a Lagrange multiplier
for this constraint.
This condition reads also:
\[
 0 = \bfeta_{a;j} + [t_j,\bfeta_a] = g^{-1}\left(g\bfeta_ag^{-1}\right)_{;j}g.
\]
It is equivalent to say that there exists $\g$-valued functions $\bfA_a$, for $a=1,\cdots, n$,
which depends only on $x$ (and not on $g$) such that
\[
 \bfeta_a(x,g) = g^{-1}\bfA_a(x)g,\quad \forall x\in M,\forall g\in \G.
\]
Plugging this expression in the first equation in System (\ref{ham-volt-beau}) it has the consequence that
\[
 \textsf{h}^*\bfp_{ab} = g^{-1}\bfPhi_{ab}g,
\]
where $\bfPhi_{ab}:= \bfA_{b;a} - \bfA_{a;b} + [\bfA_a,\bfA_b]$ does not depend on
$g$. We then observe that
\begin{equation}\label{observation1}
 \left(\textsf{h}^*\bfp^{ab}\right)_{;b} + [\bfeta_b,\textsf{h}^*\bfp^{ab}] = g^{-1}\left(
 \bfPhi^{ab}_{;b} + [\bfA_b,\bfPhi^{ab}]\right)g.
\end{equation}
\item The third equation in (\ref{ham-volt-beau}) can be translated by
using (\ref{adjoint-de-ad}) to the form:
\begin{equation}\label{equation3}
\left(\textsf{h}^*\bfp^{ab}\right)_{;b} + [\bfeta_b,\textsf{h}^*\bfp^{ab}]
+ \left(\textsf{h}^*\bfp^{aj}\right)_{;j} + [t_j,\textsf{h}^*\bfp^{aj}]
- \hbox{tr}(\hbox{ad}_{t_j})\textsf{h}^*\bfp^{aj} = 0,
\end{equation}
Let us set $\bfPhi^{aj}:= g(\textsf{h}^*\bfp^{aj})g^{-1}$, this implies that
\begin{equation}\label{observation2}
 \left(\textsf{h}^*\bfp^{aj}\right)_{;j}  + [t_j,\textsf{h}^*\bfp^{aj}] = g^{-1}\bfPhi^{aj}_{;j}g.
\end{equation}
In view of (\ref{observation1}) and (\ref{observation2}), (\ref{equation3}) is equivalent to
\begin{equation}\label{equation-magique}
 \bfPhi^{ab}_{;b} + [\bfA_b,\bfPhi^{ab}]
 + \bfPhi^{aj}_{;j} - \hbox{tr}(\hbox{ad}_{t_j})\bfPhi^{aj} = 0.
\end{equation}
\end{enumerate}
We then have the result:
\begin{theo}
 Assume that $\g$ is unimodular and that $\G$ is compact, then for any solution to
 (\ref{ham-volt-beau}), the 1-form $\bfeta$ is a solution of the classical Yang--Mills
 equations.
\end{theo}
\emph{Proof} --- The assumption that $\g$ is unimodular leads to the simplification of
(\ref{equation-magique}):
\[
 \bfPhi^{ab}_{;b} + [\bfA_b,\bfPhi^{ab}]
 = - \bfPhi^{aj}_{;j}.
\]
We observe that the left hand side of this relation
does not depend on $g\in \G$ (because
$\bfA_a$ and hence $\bfPhi^{ab}$ are constant along the fibers of $\mathcal{P}$). Hence
the same is true for $\bfPhi^{aj}_{;j}$.

For any $x\in \mathcal{M}$, consider the restriction of the $\g$-valued
$(r-1)$-form $\bfPhi^{aj}\gamma_j$ on
the fiber $\mathcal{P}_x$. Corollary \ref{corollary-unimodular} implies that
\[
 d\left(\bfPhi^{aj}\gamma_j|_{\mathcal{P}_x}\right) = d\bfPhi^{aj}\wedge \gamma_j|_{\mathcal{P}_x} 
 = \bfPhi^{aj}_{;j}\gamma|_{\mathcal{P}_x}.
\]
Hence, since the fiber $\mathcal{P}_x$ is compact and $\bfPhi^{aj}_{;j}$ is constant on $\mathcal{P}_x$,
\[
\bfPhi^{aj}_{;j} \hbox{Vol}(\mathcal{P}_x) = 
 \bfPhi^{aj}_{;j}\int_{\mathcal{P}_x}\gamma = \int_{\mathcal{P}_x}\bfPhi^{aj}_{;j}\gamma
 = \int_{\mathcal{P}_x}d\left(\bfPhi^{aj}\gamma_j\right) = 0,
\]
thus $\bfPhi^{aj}_{;j} = 0$. Hence Equation (\ref{equation-magique}) gives us
\[
 \bfPhi^{ab}_{;b} + [\bfA_b,\bfPhi^{ab}] = 0,
\]
i.e. the Yang--Mills system.\hfill $\square$

\section{The Lagrangian action and gauge symmetries}
\subsection{The Lagrangian action}\label{sectionlagrangian}
It is easy to deduce from our Hamiltonian multisymplectic model a Lagrangian
formulation (see e.g. \cite{heleinleeds}). We first restrict the multisymplectic manifold to the level
set $H^{-1}(0)$. In coordinates this amounts to eliminate the coordinate $\epsilon$
through the relation $\epsilon = \frac{1}{4}\textsf{h}^{ij}p^{ab}_ip_{abj}$. Any
submanifold $\bfGamma$ in $H^{-1}(0)$ which is a graph over $\mathcal{P}$ of a map
$\bfu$ is then given by the collection of $\g$-valued
functions $\bfeta_a$ and of $\g^*$-valued functions $\bfp^{ab}$
and $\bfp^{aj}$. We define the value of the Lagrangian density $L$ at
$(\bfeta,\bfp) = (\bfeta_a,\bfp^{ab},\bfp^{aj})$ by
\[
 L(\bfeta,\bfp)\beta\wedge \gamma =
 \bfu^*\theta.
\]
The computation of $L(\bfeta_a,\bfp^{ab},\bfp^{aj})$ is relatively easy:
one deduces from (\ref{deta+eta-square}) that
$\bfu^*(d\eta+\eta\wedge \eta) = \frac{1}{2}(\bfeta_{b;a}- \bfeta_{a;b} + [\bfeta_a,\bfeta_b])\beta^a\wedge \beta^b
- (\bfeta_{a;j}+[t_j,\bfeta_a])\beta^a\wedge \gamma^j$ and obviously we have
$\bfu^*(\epsilon\beta\wedge\gamma) = \epsilon\beta\wedge\gamma$ and
$\bfu^*p = -\frac{1}{2}\bfp^{ab}\beta_{ab}\wedge \gamma + (-1)^n\bfp^{aj}\beta_a\wedge \gamma_j$.
A straightforward computation thus gives us:
\begin{equation}\label{new-Lagrangian}
 L(\bfeta,\bfp)
 = \frac{1}{4}\textsf{h}^{ij}\bfp^{ab}_i\bfp_{abj}
 -\frac{1}{2}\bfp^{ab}(\bfeta_{b;a} - \bfeta_{a;b} + [\bfeta_a,\bfeta_b])
 + \bfp^{aj}(\bfeta_{a;j} + [t_j,\bfeta_a]).
\end{equation}
Critical points of the functional $\int_\mathcal{P} L(\bfeta_a,\bfp^{ab},\bfp^{aj})\beta \wedge \gamma$ are
the solutions of the HVDW system of equations (\ref{ham-volt-beau}).

Alternatively we may decompose $\bfeta = g^{-1}dg + g^{-1}\bfA g$ as in
(\ref{variant-normalization-in-trivialisation}) and replace the dual variables
$\bfp$ by the $\g$-valued $(n+r-2)$-form
$\bfPhi$ such that $\textsf{h}^*\bfp = g^{-1}\bfPhi g$. Then our action functional reads
\[
 L(\bfA,\bfPhi) = \frac{1}{4}h\left(\bfPhi^{ab},\bfPhi_{ab}\right)
 - \frac{1}{2}\textsf{h}_*\bfPhi^{ab}\left(\bfA_{b;a} - \bfA_{a;b} + [\bfA_a,\bfA_b]\right)
 + \textsf{h}_*\bfPhi^{aj}\bfA_{a;j}.
\]

\subsection{Gauge symmetries}
Our variational problem is invariant under the action of the gauge group of the standard Yang--Mills
action. We set this gauge group to be:
\[
\mathcal{G}:= \{\bfgamma:\mathcal{P}\longrightarrow \G; \forall z\in \mathcal{P},\forall g\in \G,
\bfgamma(z\cdot g) = g^{-1}\bfgamma(z)g\}.
\]
Note that, through a local trivialization of $\mathcal{P}$ induced by a section $\bfsigma:\mathcal{M}\longrightarrow \mathcal{P}$,
we can represent all maps $\bfgamma\in \mathcal{G}$ in the form
\begin{equation}\label{gauge-transform-generator}
 \bfgamma(z) = \bfgamma(\bfsigma(x)\cdot g) = g^{-1}\bff(x)g, 
\end{equation}
where $\bff:\mathcal{M}\longrightarrow \G$ is an arbitrary map. The gauge group $\mathcal{G}$
acts on $\Gamma_\textsc{n}(\mathcal{P},\g\otimes T^*\mathcal{P})$ through the transformation
\begin{equation}\label{gauge-transform-level1}
 \bfeta \longmapsto \widetilde{\bfeta}:= \bfgamma^{-1}d\bfgamma + \bfgamma^{-1}\bfeta \bfgamma.
\end{equation}
Indeed in the decomposition $\bfeta = g^{-1}dg  + \bfeta_a\beta^a$, we compute that
\[
 \widetilde{\bfeta} = g^{-1}dg + \left[g^{-1}(\bff^{-1}d\bff)g + \bfgamma^{-1}\bfeta_a\beta^a\bfgamma\right],
\]
confirming that $\widetilde{\bfeta}$ is still normalized. Alternatively if we 
write $\bfeta = g^{-1}dg + g^{-1}\bfA g$, we then obtain $\widetilde{\bfeta} =
g^{-1}dg + g^{-1}\widetilde{\bfA} g$, where $\widetilde{\bfA}:= \bff^{-1}d\bff + \bff^{-1}\bfA \bff$.
This shows also that, if $\bfeta$ is \emph{normalized} and \emph{equivariant}, i.e. if $\bfA$ does not depend on $g\in \G$,
then $\widetilde{\bfeta}$ is also normalized and equivariant.
We also observe that
\begin{equation}\label{gauge-transform-level2}
 d\widetilde{\bfeta} + \widetilde{\bfeta}\wedge \widetilde{\bfeta} =
 \bfgamma^{-1}(d\bfeta + \bfeta\wedge \bfeta)\bfgamma
 = \hbox{Ad}_{\bfgamma^{-1}}(d\bfeta + \bfeta\wedge \bfeta).
\end{equation}
We extend this action of $\mathcal{G}$ on sections of
$\R\oplus_\mathcal{P} \left(\g\otimes^\textsc{n}T^*\mathcal{P}\right)\oplus_\mathcal{P}\left(\g^*\otimes \Lambda^{n+r-2}T^*\mathcal{P}\right)$
over $\mathcal{P}$ by letting
\begin{equation}\label{gauge-transform-level3}
 \bfp \longmapsto \widetilde{\bfp}:= \hbox{Ad}_{\bfgamma}^*\bfp.
\end{equation}
Then $\bfp\wedge (d\bfeta + \bfeta\wedge \bfeta)$ is transformed
as follows
\[
 \begin{array}{ccl}
  \bfp\wedge (d\bfeta + \bfeta\wedge \bfeta) & \longmapsto & 
  \hbox{Ad}_{\bfgamma}^*\bfp \wedge \hbox{Ad}_{\bfgamma^{-1}}(d\bfeta + \bfeta\wedge \bfeta)\\
  & = & \bfp\wedge [\hbox{Ad}_{\bfgamma}\circ \hbox{Ad}_{\bfgamma^{-1}}(d\bfeta + \bfeta\wedge \bfeta)]
  = \bfp\wedge (\bfeta + \bfeta\wedge \bfeta),
 \end{array}
\]
i.e. is invariant by the gauge action. Hence $\theta$ is invariant by the gauge action and,
obviously the Hamiltonian function $H$ also.

\subsection{An alternative action of the gauge group}
The gauge group $\mathcal{G}$ has a different action on $\Gamma_\textsc{n}(\mathcal{P},\g\otimes T^*\mathcal{P})$.
First we observe that any $\bfgamma\in \mathcal{G}$ acts on $\mathcal{P}$ by the map $\varphi:z\longmapsto z\cdot \bfgamma(z)$.
This induces the action by pull-back $\bfeta \longmapsto \varphi^*\bfeta$
on sections of $\g\otimes T^*\mathcal{P}$.
If $\bfeta$ is normalized and has the form $\bfeta_{(x,g)} = g^{-1}dg + \bfeta_a(x,g)\beta^a$ in a local trivialization, then
$\varphi^*\bfeta_{(x,g)} = g^{-1}dg + [\bfeta_a(x,\bff(x)g)\beta^a + g^{-1}(\bff^{-1}d\bff)g]$, which
shows in particular that $\varphi^*\bfeta$ is still normalized.
If furthermore $\bfeta$ is equivariant and reads $\bfeta = g^{-1}dg + g^{-1}\bfA_a(x)\beta^a g$, then
$\varphi^*\bfeta = g^{-1}dg + g^{-1}\widetilde{\bfA}_a(x)\beta^ag$, where
$\widetilde{\bfA}_a = \bff^{-1}d\bff + \bff^{-1}\bfA_a\bff$.
Hence this action coincides with the previous one on the \emph{equivariant} normalized sections
of $\g\otimes T^*\mathcal{P}$. However it differs from the previous one on non equivariant nomalized
sections. In particular the Lagrangian given by (\ref{new-Lagrangian}) is not invariant off-shell by this
gauge action. It
is  however an on-shell symmetry if $\G$ is unimodular and compact, since then any solution of the
HVDW is equivariant.

\subsection{Gauge symmetries on dual fields}
Our action functional is also invariant under the action of another group, which is additive
(and hence Abelian).
This group is parametrized by the space $\mathcal{G}^\star$ of sections
$\bfU$ of the bundle $\g^*\otimes_\mathcal{P}\pi_\mathcal{M}^*T\mathcal{M}\otimes_\mathcal{P}\Lambda^{r-1}T^*\mathcal{P}$ over $\mathcal{P}$ and which satisfy
\begin{equation}\label{exoticgaugegroup}
 \left(d\bfU - \hbox{ad}_\alpha^*\wedge \bfU\right)|_{\mathcal{P}_x} = 0,
 \quad \forall x\in \mathcal{M},
\end{equation}
where $\alpha$ is given by (\ref{definition-maurer-cartan-droite}).
This definition requires some comments: for any $z\in \mathcal{P}$, the value of $\bfU$ at
$z$ is a $(r-1)$-form with coefficients in
$\g^*\otimes T_x\mathcal{M}$, where $x=\pi_\mathcal{M}(z)$, hence we can write $\bfU = \bfU^a\underline{e}_a$, where
$(\underline{e}_1,\cdots,\underline{e}_n)$ is a basis of $T_x\mathcal{M}$ and each $\bfU^a$ is a $\g^*$-valued
$(r-1)$-form. Then Condition (\ref{exoticgaugegroup}) means that
$\left(d\bfU^a - \gamma^i\wedge \hbox{ad}_{t_i}^*\bfU^a\right)|_{\mathcal{P}_x} = 0$, for any $a$.
If we set $\bfU^a = \textsf{h}_*\bfpsi^a\Longleftrightarrow \bfpsi^a = \textsf{h}^*\bfU^a$, where $\bfpsi^a$ is a
$\g$-valued $(r-1)$-form, then the latter condition reads
$\left(d\bfpsi^a + [g^{-1}dg,\bfpsi^a]\right)|_{\mathcal{P}_x} = 0$ or equivalently
$\beta\wedge \left(d\bfpsi^a + [g^{-1}dg,\bfpsi^a]\right) = 0$. Solutions $\bfpsi^a$ of this equation
are of the form $\bfpsi^a = g^{-1}\bfvarphi^a g$, where $\bfvarphi^a\in \g\otimes\Omega^{r-2}\mathcal{P}$
is \emph{closed}.
In conclusion $\bfU = \underline{e}_a\textsf{h}_*\left(\hbox{Ad}_{g^{-1}}\bfvarphi^a\right)$,
where $d\bfvarphi^a = 0$.

The action of any $\bfU\in \mathcal{G}^\star$ is defined by
$(\eta,p)\longmapsto (\eta,p + (-1)^n\beta_a\wedge \bfU^a)$. Since components $p^{ab}$
are left unchanged, the Hamiltonian function $H$ is obviously invariant. Moreover
under this gauge action $\theta$ is changed into
\[
 \theta + \beta\wedge \left(d\eta_a+[g^{-1}dg,\eta_a]\right)\wedge \bfU^a
 = \theta + (-1)^nd\left(\beta\wedge \hbox{Ad}_g\eta_a\wedge \textsf{h}_*\bfvarphi^a\right),
\]
so that we see that $\theta$ is affected by the addition of an exact form and, in particular, $\omega = d\theta$
is left unchanged.
An alternative description of this gauge group is that it coincides with
sections $\bfV$ of $\g^*\otimes \Lambda^{n+r-2}_1T^*\mathcal{P}$ mod $\g^*\otimes \Lambda^{n+r-2}_0T^*\mathcal{P}$
(see Paragraph \ref{reinterpretation}) which satisfy $d\bfV - \hbox{ad}_\alpha^*\wedge \bfV = 0$,
since any such section has the form
$\beta_a\wedge \bfU^a$, where $\bfU\in \mathcal{G}^\star$.

Using the variables $\bfPhi$ as in  Paragraph \ref{sectionlagrangian},
the $\mathcal{G}^\star$ gauge action reads $(\bfA,\bfPhi)\longmapsto (\bfA,\bfPhi + \bfchi)$,
where $\bfchi^{ab} = 0$ and $\bfchi^{aj}$ satisfies $\bfchi^{aj}_{;j} = 0$ or
equivalently $d(\bfchi^{aj}\gamma_j) = 0$.

\subsection{Gauge fixing}
We can fix the action of $\mathcal{G}$ by choosing a critical point (with respect to $\mathcal{G}$ deformations)
of the functional
$\int_\mathcal{P} \frac{1}{2}h(\bfeta_a,\bfeta^a)\beta\wedge \gamma
=\int_\mathcal{P} \frac{1}{2}h(\bfA_a,\bfA^a)\beta\wedge \gamma$. It leads to the condition
$\int_{\mathcal{P}_x}\hbox{Ad}_g(\bfeta^a_{;a})\gamma = \int_{\mathcal{P}_x}\bfA^a_{;a}\gamma = 0$, $\forall x\in \mathcal{M}$.

Similarly the action of $\mathcal{G}^*$ can fixed by using, for each $x\in \mathcal{M}$, a Hodge decomposition
of the $\g$-valued $(r-1)$-form $\bfPhi^{aj}\gamma_j|_{\mathcal{P}_x}$. This leads to choose $\bfPhi^{aj}$ such that,
for any $x\in \mathcal{M}$, $\forall a$,
$\bfPhi^{aj}\gamma_j|_{\mathcal{P}_x} = \bf\textsf{h}^a +  *d\bfV^a$, where $\bfV^a$ is a function
from $\mathcal{P}_x$ to $\g$ and $\bfh^a|_{\mathcal{P}_x}$ is
a harmonic $\g$-valued $(r-1)$-form on $\mathcal{P}_x$ (note that $\bfh^a = 0$ if
the de Rham cohomology group $H^{r-1}(\G)$ is trivial).

Putting these gauge fixing conditions together with equations (\ref{ham-volt-beau}) then leads to
a well-posed system, which, if $H^{r-1}(\G) = \{0\}$, reduces to the standard Yang--Mills system in the Lorentz gauge.


\begin{thebibliography}{100}

\bibitem{Baez} J. C. Baez, C. L. Rogers, \emph{Categorified Symplectic Geometry and the String Lie 2-algebra}, 
Homology, Homotopy and Applications, Vol. 12, N. 1 (2010), 221--236.
%preprint arXiv:0901.4721


%\bibitem{brouder} C. Brouder, \emph{Runge--Kutta methods and renormalization}, Eur. Phys. J. C. 12 (2000), 521--534.

%\bibitem{brouder2} C. Brouder, \emph{Trees, renormalization and differential equations}, B.I.T. 44 (2004), no. 6, 425--438.

%\bibitem{butcher} J.C. Butcher, \emph{The numerical analysis of ordinary differential equations}, Wiley, Chichester, 1987.

\bibitem{BrunoCV} D. Bruno, R. Cianci, S. Vignolo, \emph{On the Hamiltonian formulation of Yang–Mills gauge theories},
International Journal of Geometric Methods in Modern Physics, Vol. 2, Issue 06 (2005/12), 1115--1131.
%arXiv:math-ph/0507001.


\bibitem{CantrijnIdL} F. Cantrijn, A. Ibort, and M. de Leon, \emph{On the geometry of multisymplectic manifolds},
J. Austral. Math. Soc. Ser. A, 66(3):303–330, 1999.


\bibitem{caratheodory} C. Carath{\'e}odory, {\em Variationsrechnung und partielle
Differentialgleichungen erster Ordnung}, Teubner, Leipzig
(reprinted by Chelsea, New York, 1982); Acta litt. ac scient. univers. Hungaricae, Szeged, Sect. Math., 4 (1929), p. 193.

\bibitem{cartan} E. Cartan, \emph{Le{\c c}ons sur les invariants int{\'e}graux}, Hermann, 1922.

%\bibitem{chen} K.T. Chen, \emph{Integration of paths, geometric invariants and a generalized Baker-Hausdorff formula}, Annals of Math. 65 (1957), 163--178.

%\bibitem{clebsch} A. Clebsch, \emph{Ueber die zweite Variation vielfache Integralen}, J. reine angew. Math. 56 (1859), 122--148.

%\bibitem{CrnkovicWitten} C. Crnkovic, E. Witten, \emph{Covariant description of canonical formalism in geometrical theories}, in \emph{Three hundred years of gravitation}, 676--684; E. Witten, \emph{Interacting field theory of open supertrings}, Nucl. Phys. B276, 291.

\bibitem{Dedecker} P. Dedecker, {\em Calcul des variations, formes diff{\'e}rentielles et
champs g{\'e}od{\'e}siques}, in {\em G{\'e}om{\'e}trie diff{\'e}rentielle}, Colloq. Intern. du CNRS LII,
Strasbourg 1953, Publ. du CNRS, Paris, 1953, p. 17-34; {\em On the generalization of
symplectic geometry to multiple integrals in the calculus of variations}, in {\em Differential
Geometrical Methods in Mathematical Physics}, eds. K. Bleuler and A. Reetz, Lect. Notes Maths.
vol. 570, Springer-Verlag, Berlin, 1977, p. 395-456.

%\bibitem{oldDD}  T. De Donder, {\em Introduction {\`a} la th{\'e}orie des invariants int{\'e}graux}, Bull. Acad. Roy. Belgique 1913, 1043--1073.
%Voir aussi dans les Rend. Circ. Mat. Palermo t. XV (1901), 66--131 : Etude sur les invariants int\'egraux;\\ et dans le Bull. Acad. Roy. Belgique : 1908, 795--811; 1909, 66-83; 1909, 610--621; 1909, 268--286, 1911, 50--70; 1911, 740--759, 1913, 1043--1073\\ Th\'eorie invariantive du calcul des variations, 1929, chap. 1 + une note de Lepage dans le livre de m{\^e}me titre de 1930.

\bibitem{DeDonder} T. De Donder, {\em Th{\'e}orie invariantive du calcul des variations},
Gauthiers-Villars, Paris, 1930.

%\bibitem{DolanHaugh} B.P. Dolan, K.P. Haugh, \emph{A co-variant approach to Ashtekar's canonical gravity}, Class. Quant. Gravity, Vol. 14, N. 2, 1997,  477--488 (12).

%\bibitem{duetsch} M. D{\"u}tsch, K. Fredenhagen, \emph{The master Ward identity and generalized Schwinger--Dyson equation in classical field theory}, Commun. Math. Phys. (2003)

\bibitem{forgerromero} M. Forger, S. V. Romero, \emph{Covariant Poisson bracket in geometric field theory}, Commun. Math. Phys. 256 (2005), 375--410.

\bibitem{forgergomes} M. Forger, L. Gomes, \emph{Multisymplectic and polysymplectic structures on fiber bundles},
Rev. Math. Phys. 25 (2013), no. 9.
%\bibitem{kijowski} J. Kijowski, \emph{A finite dimensional canonical formalism in the classical field theory}, Comm. Math. Phys. 30 (1973), 99-128.

%\bibitem{kreimer} D. Kreimer, \emph{ On the Hopf algebra structure of perturbative quantum field theory}, Adv. Th. Math. Phys., 2 (1998), pp. 303--334.

\bibitem{garcia} P. L. Garc{\'\i}a, \emph{Geometr{\'\i}a simpl{\'e}tica en la teoria de campos}, Collect. Math. 19, 1--2, 73, 1968.

%\bibitem{garcia-perez-rendon}P. L. Garc{\'\i}a, A. P{\'e}rez-Rend{\'o}n, \emph{Symplectic approach to the theory of quantized fields, I}, Commun. Math. Phys. 13 (1969), 24--44 and \emph{---, II}, Arch. Rational Mech. Anal.  43  (1971), 101--124. 

\bibitem{GIMMSY} M.J. Gotay, J. Isenberg, J.E. Marsden (with the collaboraton of R. Montgomery,
J. \'Snyatycki, P.B. Yasskin),
{\em Momentum maps and classical relativistic fields, Part I/ covariant
field theory}, preprint arXiv/physics/9801019

%\bibitem{Gawedski} K. Gaw\c{e}dski, {\em On the generalization of the canonical formalism in the classical field theory}, Rep. Math. Phys. No 4, Vol. 3 (1972), 307--326.

\bibitem{goldstern} H. Goldschmidt, S. Sternberg, \emph{The Hamilton--Cartan formalism in the calculus of variations}, Ann. Inst. Fourier Grenoble 23, 1 (1973), 203--267.

%\bibitem{Hadamard} J. Hadamard, \emph{Sur une question du calcul des variations}, Bull. Soc. Math. France 30 (1902), 253--256; \emph{Sur quelques questions du calcul des variations}, ibid. 33 (1905), 73--80..

%\bibitem{hairerwanner} E. Hairer and G. Wanner, \emph{On the {Butcher} group and general multi--value methods}, Computing (1974), 13 (1), 1--15.

%\bibitem{harrivel} D. Harrivel, \emph{Planar binary trees and perturbative calculus of observables in classical field theory}, Ann. IHP 23 (2006), 891--909.

%\bibitem{harrivel2} D. Harrivel, \emph{Butcher series and control theory}, preprint arXiv math/0603133

%\bibitem{harrivelhelein} D. Harrivel, F. H{\'e}lein, \emph{Geometric covariant quantization of fields}, in preparation.

\bibitem{hk1} F. H{\'e}lein, J. Kouneiher, \emph{Covariant Hamiltonian formalism for the calculus of variations with several
variables: Lepage--Dedecker versus De Donder--Weyl}, Adv. Theor. Math. Phys. 8 (2004), 565--601.

\bibitem{hk2} F. H{\'e}lein, J. Kouneiher, \emph{The notion of observable in the covariant Hamiltonian
formalism for the calculus of variations with several variables}, Adv. Theor. Math. Phys. 8 (2004), 735--777.

\bibitem{helein} F. H{\'e}lein, \emph{Hamiltonian formalisms for multidimensional calculus of variations and perturbation theory}, in \emph{Noncompact problems at the intersection of geometry, analysis, and topology}, Contemp. Math., 350 (2004),  127--147.

\bibitem{heleinleeds} F. H{\'e}lein, \emph{Multisymplectic formalism and the covariant phase space},
in \emph{Variational Problems in Differential Geometry}, Roger Bielawski, Kevin Houston, Martin Speight, eds,
London Mathematical Society Lecture Note Series 394, Cambridge University Press, 2012, p. 94-126.

%\bibitem{heleinprep} F. H{\'e}lein, \emph{The use of the covariant phase space on non nonlinear fields}, in preparation.


%\bibitem{jacobi-hamilton} C.G.J. Jacobi, \emph{Ueber die Reduction der Integration des partiellen Differentialgleichungen erster Ornung zwischen irgend einer Zahl Variabeln auf die Integration eines einzigen Systemes gew{\"o}hnlicher Differentialgleichungen}, J. reine angew. Math. 17 (1837), 68--82.

%\bibitem{jacobi-legendre} C.G.J. Jacobi, \emph{Zur Theorie des Variations-Rechnung und des Differential-Gleichungen}, J. reine angew. Math. 17 (1837), 97--162.

\bibitem{kanatchikov1}
I. V. Kanatchikov, {\em Canonical structure of classical field theory in the polymomentum
phase space}, Rep. Math. Phys. vol. 41, No. 1 (1998).
%arXiv:hep-th/9709229


\bibitem{kanatchikov1.5}
I. V. Kanatchikov, {\em Precanonical quantization of Yang-Mills fields and the functional Schrödinger representation},
Rep. Math. Phys. 53 (2004), 181--193; hep-th/0301001



\bibitem{kanatchikov2}
I. V. Kanatchikov, \emph{On the precanonical structure of the Schrödinger wave functional}, arXiv:1312.4518

\bibitem{Kijowski1} J. Kijowski, {\em A finite dimensional canonical formalism in the classical
field theory}, Comm. Math. Phys. 30 (1973), 99-128.

\bibitem{Kijowski2} J. Kijowski, {\em Multiphase spaces and gauge in the
calculus of variations}, Bull. de l'Acad. Polon. des Sci., S{\'e}rie sci. Math., Astr. et Phys. XXII (1974), 1219-1225.

\bibitem{KijowskiSzczyrba} J. Kijowski, W. Szczyrba, \emph{A canonical structure for classical field theories}, Commun. Math Phys. 46 (1976), 183--206.

%\bibitem{kirillov} A.A. Kirillov, \emph{Geometric quantization}, in \emph{Dynamical systems IV}, V.I. Arnol'd, S.P. Novikov, eds., Springer-Verlag, 1990.

%\bibitem{koenigsberger} L. Koenigsberger, \emph{Die Prinzipien der Mechanik f{\"u}r mehrere Variable}, Sitzungsberichte Akad. Wiss. Berlin, Bd. XLVI, 14 nov. 1901, 1108; \emph{Die Prinzipien der Mechanik f{\"u}r mehrere unavh{\"a}ngige Variable}, J. Reine Angw. Math., Bd. 124 (1902), 202--277.

\bibitem{krupka} D. Krupka, \emph{A geometric theory of ordinary first order variational problems
in fibered manifolds}, I. \emph{Critical sections}, J. Math. Anal. Appl. 49 (1975), 180--206;
II. \emph{Invariance}, J. Math. Anal. Appl. 49 (1975),469--476.

%\bibitem{yks} Y. Kosmann-Schwarzbach, \emph{Les th{\'e}or{\`e}mes de Noether ---Invariance et lois de conservation au XXe si{\`e}cle}, Les {\'e}ditions de l'Ecole Polytechnique, 2004.

\bibitem{Lepage} T. Lepage, \emph{Sur les champs g{\'e}od{\'e}siques du calcul des variations}, Bull. Acad. Roy. Belg., Cl. Sci. 27 (1936), 716--729, 1036--1046.

\bibitem{lopez-marsden} M.C. L{\'o}pez and J.E. Marsden, \emph{Some remarks on Lagrangian and Poisson reduction for ﬁeld theories}, J. Geom. Phys. 48 (2003) 52-83.

%\bibitem{Noether} E. Noether, \emph{Invariante Variationsprobleme}, Nachrichten von der K{\"o}niglichen Gesellschaft des Wissenschaften su G{\"o}ttingen, Mathematisch-physikalische Kalsse, 1918, p. 235--257.

%\bibitem{peierls} R.E. Peierls,\emph{The commutation laws of relativistic field theory}, Proc. Roy. Soc. London, Ser. A, Vol. 214, No. 1117 (1952), 143--157.

%\bibitem{prange} G. Prange, \emph{Die Hamilton--Jacobische Theorie f{\"u}r Doppelintegrale}, Diss. G{\"o}ttingen, 1915.

%\bibitem{poincare} H. Poincar{\'e}, \emph{Les m{\'e}thodes nouvelles de la m{\'e}canique c{\'e}leste}, t. III, Paris, Gauthier--Villars, 1899.

%\bibitem{reyes} E.G. Reyes, \emph{On covariant phase space and the variational bicomplex}, Int. J. Theor. Phys., Vol. 43, No. 5, 2004.

\bibitem{Rovelli} C. Rovelli, {\em A note on the foundation of relativistic mechanics ---
II: Covariant Hamiltonian general relativity}, arXiv:gr-qc/0202079

%\bibitem{Rund} H. Rund, \emph{The Hamilton--Jacobi theory in the calculus of variations, its role in mathematics and physics}, Krieger Pub. 1973.

%\bibitem{segal} I. Segal, \emph{Quantization of nonlinear systems}, J. Math. Phys. vol. 1, N. 6 (1960), 468--488.

\bibitem{snyatycki} J. {\'S}nyatycki, \emph{Geometric quantization and quantum Mechanics}, Appl. Math. Sci. 30, Springer-Verlag 1980.

%\bibitem{souriau} J.-M. Souriau, \emph{Structure des syst{\`e}mes dynamiques}, Dunod, Paris, 1970.

%\bibitem{takens} F. Takens, \emph{A global formulation of the inverse problem of the calculus of variations}, J. Diff. Geom. 14 (1979), 543--562.

%\bibitem{VanHove} L. Van Hove, \emph{Sur les champs de De Donder--Weyl et leur construction par la méthode des caractéristiques}, Bull. Acad. R. Belg. 31 (1945), 278--285; \emph{Sur les champs de Carathéodory et leur construction par la méthode des caractéristiques}, ibid., 625--638.

%\bibitem{vinogradov} A.M. Vinogradov, \emph{The $\mathcal{C}$-spectral sequence, Lagrangian formalism, and conservations laws, I and II}, J. Math. Anal. Appl. 100 (1984), 1--40 and 41--129.

%\bibitem{Vitagliano} L. Vitagliano, \emph{Secondary calculus and the covariant phase space}, preprint diffiety.org

%\bibitem{vita1} L. Vitagliano, \emph{The Lagrangian--Hamiltonian Formalism for Higher Order Field Theories }, preprint arXiv:0905.4580

%\bibitem{vita2} L. Vitagliano, \emph{Partial Differential Hamiltonian Systems}, preprint arXiv:0903.4528

\bibitem{volterra1} V. Volterra, \emph{Sulle equazioni differenziali che provengono da questiono di calcolo delle variazioni}, Rend. Cont. Acad. Lincei, ser. IV, vol. VI, 1890, 42--54.

\bibitem{volterra2} V. Volterra, \emph{Sopra una estensione della teoria Jacobi--Hamilton del calcolo delle varizioni},  Rend. Cont. Acad. Lincei, ser. IV, vol. VI, 1890, 127--138.

\bibitem{Weyl} H. Weyl, {\em Geodesic fields in the calculus of variation for multiple
integrals}, Ann. Math. (3) 36 (1935), 607--629.

%\bibitem{zuckerman} G. Zuckerman, \emph{Action functional and global geometry}, in \emph{Mathematical aspects of string theory}, S.T. Yau, eds., Advanced Series in Mathematical Physics, vol 1, World Scientific, 1987, 259--284.

\end{thebibliography}
\end{document}